\documentclass[12pt]{article}
\usepackage{jheppub}

\pdfoutput=1

\usepackage{amsmath,bbm,array,amsfonts,graphicx,wrapfig,lscape,float,mathtools,multirow,longtable}
\usepackage[dvipsnames]{xcolor}
\usepackage{array}

\newcommand{\be}{\begin{equation}}
\newcommand{\ee}{\end{equation}}
\newcommand{\beq}{\begin{equation}}
\newcommand{\beql}[1]{\begin{equation}\label{#1}}
\newcommand{\eeq}{\end{equation}}
\newcommand{\ba}{\begin{array}}
\newcommand{\ea}{\end{array}}
\newcommand{\bea}{\begin{eqnarray}}
\newcommand{\beal}[1]{\begin{eqnarray}\label{#1}}
\newcommand{\eea}{\end{eqnarray}}
\newcommand{\ben}{\begin{enumerate}}
\newcommand{\een}{\end{enumerate}}
\newcommand{\bean}{\begin{eqnarray*}}
\newcommand{\eean}{\end{eqnarray*}}
\newcommand{\eref}[1]{(\ref{#1})}
\newcommand{\sref}[1]{\S\ref{#1}}

\newcommand{\fref}[1]{Figure \ref{#1}}
\newcommand{\btab}[1]{\begin{tabular}{#1}}
\newcommand{\etab}{\end{tabular}}

\def \Black {\pdfsetcolor{0 0 0 1}}

\newcommand{\comment}[1]{}

\newcommand{\qed}{\nobreak \ifvmode \relax \else
      \ifdim\lastskip<1.5em \hskip-\lastskip
      \hskip1.5em plus0em minus0.5em \fi \nobreak
      \vrule height0.75em width0.5em depth0.25em\fi}

\definecolor{darkspringgreen}{rgb}{0.09, 0.45, 0.27}
\definecolor{forestgreen}{rgb}{0.13, 0.55, 0.13}

\usepackage{array}
\usepackage{physics}
\usepackage{subcaption}
\usepackage{tikz,tikz-3dplot}
\usepackage[colorlinks=true]{hyperref}
\newcolumntype{C}[1]{>{\centering\let\newline\\\arraybackslash\hspace{0pt}}m{#1}}

\definecolor{yellow2}{rgb}{0.98, 0.80, 0.20}

\title{Quiver Tails and Brane Webs}

\author[a,b,c]{Sebasti\'an Franco,}

\author[d,e]{Diego Rodr\'iguez-G\'omez}

\affiliation[a]{Physics Department, The City College of the CUNY\\
	160 Convent Avenue, New York, NY 10031, USA}
\affiliation[b]{Physics Program and \textsuperscript{$c$}Initiative for the Theoretical Sciences\\
	The Graduate School and University Center, The City University of New York\\
	365 Fifth Avenue, New York NY 10016, USA}
\affiliation[d]{Department of Physics, Universidad de Oviedo \\  
C/ Federico Garc\'ia Lorca  18, 33007  Oviedo, Spain}
\affiliation[e]{Instituto Universitario de Ciencias y Tecnolog\'ias Espaciales de Asturias (ICTEA) \\
 C/~de la Independencia 13, 33004 Oviedo, Spain.}
	
\emailAdd{sfranco@ccny.cuny.edu}
\emailAdd{d.rodriguez.gomez@uniovi.es}

\abstract{A new type of quiver theories, denoted twin quivers, was recently introduced for studying $5d$ SCFTs engineered by webs of 5-branes ending on 7-branes. Twin quivers provide an alternative perspective on various aspects of such webs, including Hanany-Witten moves and the $s$-rule. More ambitiously, they can be regarded as a first step towards the construction of combinatorial objects, generalizing brane tilings, encoding the corresponding BPS quivers. This paper continues the investigation of twin quivers, focusing on their non-uniqueness, which stems from the multiplicity of toric phases for a given toric Calabi-Yau 3-fold. We find that the different twin quivers are necessary for describing what we call quiver tails, which in turn correspond to certain sub-configurations in the webs. More generally, the multiplicity of twin quivers captures the roots of the Higgs branch in the extended Coulomb branch of $5d$ theories.}

\begin{document}

\maketitle

\section{Introduction} 

Quantum Field Theories in $5d$ are notoriously hard to construct. In the case of supersymmetric theories, String/M Theory provides various approaches to engineer $5d$ Superconformal Field Theories (SCFTs). For instance, M-theory on a toric Calabi-Yau 3-fold (CY$_3$) engineers a $5d$ SCFT in the transverse directions \cite{Morrison:1996xf,Intriligator:1997pq}.\footnote{The geometric engineering of $5d$ theories has been thoroughly studied beyond the toric case. See for instance \cite{DelZotto:2017pti,Xie:2017pfl,Jefferson:2017ahm,Jefferson:2018irk,Bhardwaj:2018yhy,Bhardwaj:2018vuu,Apruzzi:2018nre,Closset:2018bjz,Bhardwaj:2019jtr,Apruzzi:2019vpe,Apruzzi:2019opn,Apruzzi:2019enx,Bhardwaj:2019xeg,Saxena:2020ltf,Apruzzi:2019kgb}.} An alternative realization of $5d$ SCFTs is on the worldvolume of systems of webs of $(p,q)$ 5-branes in type IIB String Theory \cite{Aharony:1997ju,Aharony:1997bh}. In fact, it turns out that these two approaches are related by dualities, and the $(p,q)$-web corresponds to the spine of the toric diagram for the CY$_3$ \cite{Aharony:1997bh,Leung:1997tw}.

An interesting piece of information about a $5d$ SCFT is its spectrum of BPS particles, which can be encoded in a {\it BPS quiver}.\footnote{See \cite{Alim:2011kw} for an introduction to BPS quivers.} It turns out that the BPS quiver for a $5d$ theory engineered via M-theory on a toric CY$_3$ coincides with the quiver on the worldvolume of D3-branes probing the same geometry \cite{Closset:2019juk}.\footnote{To be precise, the BPS quiver is for the $5d$ theory on $S^1\times \mathbb{R}^{1,3}$.} The latter is a problem with a long and honored history that culminated with the introduction of {\it brane tilings} (also known as dimer models), which significantly simplify the connection between geometry and quiver theories \cite{Franco:2005rj,Franco:2005sm}. Brane tilings can be represented by bipartite graphs on a 2-torus that encode the quiver and superpotential of the corresponding gauge theories and, simultaneously, reduce the determination of the associated CY$_3$ to a combinatorial problem. They are physical brane configurations, connected via T-duality to the D3-branes on toric CY$_3$'s.

Brane tilings have been generalized to {\it bipartite field theories} (BFTs), a class of quiver theories defined by bipartite graphs on Riemann surfaces, which enjoy many of the combinatorial properties and connections to toric geometry of brane tilings \cite{Franco:2012mm,Franco:2012wv,Franco:2013pg,Franco:2014nca,Franco:2018vqd}.\footnote{A closely related class of theories was considered in \cite{Xie:2012mr}.} In the context of BFTs, there is an interesting operation known as {\it untwisting}, which maps a bipartite graph on Riemann surface to a new embedding of the same graph on a generically different Riemann surface. For the brane tiling encoding the BPS quiver for a $5d$ SCFT engineered by M-theory on a toric CY$_3$, the BFT obtained by untwisting is closely related to the $(p,q)$-web that engineers it. This connection is related to mirror symmetry and is another manifestation of the duality connecting both String/M Theory realizations of the $5d$ SCFT.

To visualize the Higgs branch of $5d$ theories, it is useful to terminate the legs of $(p,q)$-webs on suitable 7-branes. This opens up the possibility of having more than one 5-brane terminating on a given 7-brane. This imposes an extra constraint due to the $s$-rule, which has far reaching consequences. In these cases, the $(p,q)$-web can be regarded as the spine of a generalization of a toric diagram in which boundary edges are merged. The resulting polytopes, which include white and black dots, are called {\it generalized toric polygons} (GTPs) \cite{Benini:2009gi,vanBeest:2020kou,VanBeest:2020kxw}. GTPs raise natural new questions, including their geometric interpretation and whether the corresponding BPS quivers are captured by some objects generalizing brane tilings.

Interesting progress along these lines was made recently in \cite{Bourget:2023wlb} and \cite{Franco:2023flw}. {\it Twin quivers}, a new type of quiver theories that provide an alternative perspective on various aspects of the webs of 5- and 7-branes associated to generic GTPs, including their mutations and the $s$-rule, were introduced in \cite{Franco:2023flw}. Various methods for the construction of twin quivers were introduced in \cite{Franco:2023flw}. Untwisting of brane tilings plays an important role in them. In view of the toric case, it is natural to expect that twin quivers provide a first step towards the BPS quivers of $5d$ SCFTs associated to GTPs and their geometric description.

This paper continues the investigation of twin quivers, focusing on the physical significance of the ones obtained from brane tilings associated to multiple toric phases of a single underlying CY$_3$. This multiplicity was already noticed in \cite{Franco:2023flw}. Remarkably, we find that the different twin quivers correspond to different chambers in the extended Coulomb branch of the $5d$ SCFT sensitive to the roots of the Higgs branch. As a consequence, this provides a method for constructing GTPs including white dots starting from standard toric theories.

This paper is organized as follows. Section \sref{sect:twin_quivers} reviews the brane web and geometric realizations of $5d$ theories, GTPs and twin quivers. Section \sref{section_twin_quivers_for_different_toric_phases} provides a first approach to the issue of twin quivers arising from multiple toric phases. Section \sref{sect:conifold/Zn} starts the study of brane webs, twin quivers and their mutations for theories associated to non-chiral $\mathbb{Z}_N$ orbifolds of the conifold. Section \sref{section_multiple_toric_phases} considers additional toric phases for the same orbifolds of the conifolds, introducing the concept of quiver tails. It also considers the decoupling of tails and connects it to the operation of node merging in the construction of twin quivers. Section \sref{section_additional_examples} presents several additional examples. Section \sref{section_tails_as_building_blocks} proposes the combination of quiver tails and gluing for generating twin quivers for general GTPs. Finally, section \label{section_conclusions} presents our conclusions and future directions.

\section{Brane webs and twin quivers for $5d$ theories}\label{sect:twin_quivers}

5d SCFTs can be engineered on webs of 5-branes in Type IIB String Theory. Part of the information on these webs can be efficiently encoded in the so-called {\it generalized toric polygons} (GTPs) \cite{Benini:2009gi,vanBeest:2020kou,VanBeest:2020kxw}. In turn, every GTP can be associated to a {\it twin quiver}, which provide, for example, an alternative perspective on the mutations and $s$-rule of the associated webs. In this section we review the engineering of $5d$ SCFTs using brane webs, and the construction of the associated GTPs and twin quivers.

\subsection{Basics on 5-brane webs for $5d$ SCFTs}

Webs of 5-branes in Type IIB String Theory provide a powerful tool for studying the dynamics of a large class of $5d$ SCFTs theories \cite{Aharony:1997bh}. The 5-branes fully extend on the $(01234)$ directions, span a line on the $(56)$ plane and sit at the point $x^7=x^8=x^9=0$ in the remaining directions. Most of the interesting physics is captured by the configuration on the $(56)$ plane, to which we will refer as the plane of the web. The slopes of 5-branes on this plane are determined by their $(p,q)$ charges, with their intersections subject to $(p,q)$ charge conservation at every vertex. The $(p,q)$ charges of every 5-brane are mutually coprime. The external $(p,q)$ 5-branes of the web, to which we will often refer as legs, are assumed to terminate on a $[p,q]$ 7-brane. Each 7-brane extends on the $(01234789)$ directions and is pointlike on the plane of the web. Multiple $(p,q)$ 5-branes can end on a  single $[p,q]$ 7-brane. In those cases, there is an extra constraint coming from supersymmetry in the form of the $s$-rule.\footnote{The $s$-rule was originally introduced in \cite{Hanany:1996ie} for D3-branes linked between D5/NS5, and it was subsequently generalized to various other contexts. In \cite{Benini:2009gi}, it was formulated for generic $(p,q)$ 5-branes ending on $[r,s]$ 7-branes (in particular, in terms of generalized toric polygons, later developed in \cite{vanBeest:2020kou}). More recently, it was discussed in in \cite{Bergman:2020myx} using the fact that the SUSY condition for 5-branes ending on 7-branes is identical to that of $(p,q)$ strings ending on $[r,s]$ 7-branes \cite{Iqbal:1998xb}.} In the context at hand, the $s$-rule states that a maximum of $|ps-qr|$ $(p,q)$ 5-branes can be supersymmetrically suspended between a $[p,q]$ 7-brane and an $(r,s)$ 5-brane \cite{Benini:2009gi}. 

It is important to bear on mind that 7-branes come with a branch cut for the axio-dilaton on the plane of the web. As it is standard, we will assume that such branch cuts point radially away from the web, without crossing it. In addition, the length of a leg is not a parameter of the low energy SCFT. Using this fact, we implicitly assume large legs, i.e. very separated 7-branes, so that the axio-dilaton is approximately constant over the scales of the web. This justifies neglecting the curvature associated to the 7-branes.

$5d$ SCFTs are intrinsically strongly coupled and isolated. They generically have a moduli space which corresponds to the possible deformations of the web which do not change the positions of the external 7-branes from which the web is suspended. In turn, the deformations which change the positions of the 7-branes correspond to relevant deformations. Such deformations trigger an RG flow to an IR effective theory, which sometimes corresponds to a standard supersymmetric gauge theory, which in $5d$ is IR free.

The moduli space is divided into the Coulomb and Higgs branches, which are captured by different deformations of the underlying brane web. On the Coulomb branch, the web is deformed inside its plane. When the $5d$ SCFT admits a deformation to a gauge theory, these moduli space directions correspond to the standard Coulomb branch, in which scalars in vector multiplets take non-zero VEVs. On the other hand, along the Higgs branch, the web is separated into consistent sub-webs which slide along the 7-branes in the $(789)$ directions. When the $5d$ SCFT admits a deformation to a gauge theory, these directions correspond to the standard Higgs branch, where scalars in hypermultiplets take VEVs. 

When the $5d$ SCFT has a deformation to a gauge theory, relevant deformations can be regarded as supersymmetric VEVs for scalars in background vector multiplets coupled to the global symmetries of the gauge theory. For this reason, the combination of the Coulomb branch and relevant deformations is often referred to as the extended Coulomb branch.

Importantly for our purposes, the Higgs branch typically only touches the (extended) Coulomb branch along some lower-dimensional manifold termed the \textit{root of the Higgs branch}. A prototypical example is the rank 1 $E_1$ theory in section \sref{E1}, where the (extended) Coulomb branch touches the Higgs branch only at the origin, and consequently the Higgs branch can only be entered at the origin of the (extended) Coulomb branch.

\subsubsection{Webs and Hanany-Witten moves}

The 5-brane web engineering a given $5d$ SCFT is not unique. To begin with, the $SL(2,\mathbb{Z})$ duality of Type IIB String Theory connects webs whose legs have charges differing by a global $SL(2,\mathbb{Z})$ transformation. Let us cyclically order the legs counterclockwise, with an arbitrary leg chosen as first. Consider two webs with legs whose charges are $\ell_i=(p_i,q_i)$ and $\ell'_i=(p'_i,q_i')$, with $i=1,\ldots,L$, respectively. The two webs define the same $5d$ SCFT if there is a matrix $U\in SL(2,\mathbb{Z})$ such that $\ell'_i=U\ell_i$ for all $i$.

More interestingly, webs can be related by crossing 7-branes. As mentioned above, since the length of a leg is not a parameter in the low energy SCFT, one may shrink it by moving the corresponding 7-brane along it until eventually crossing the rest of the web and sending it away to a very large distance in the opposite direction. In order to comply with the standard presentation discussed above, one has to rotate the branch cut of the crossed 7-brane, which, as it sweeps other 7-branes, changes them accordingly (and consequently the 5-branes). More precisely, the monodromy of a $[p_j,q_j]$ 7-brane is

\begin{equation}
M_{(p_j,q_j)}=\left(\begin{array}{cc} 1-p_j q_j & p^2 \\ -q_j^2 & 1+p_jq_j\end{array}\right)\,.
\label{monodromy}
\end{equation}
Then, when the monodromy sweeps a $[p_i,q_i]$ 7-brane counter-clockwise, it gets transformed into a $[p'_i,q'_i]^T=M_{(p_j,q_j)}\,[p_i,q_i]^T$ 7-brane (if it is swept clockwise the transformation matrix is $M^{-1}_{p_j,q_j}$). Of course, the $(p_i,q_i)$ 5-branes ending on it change accordingly. Note that in the process the crossed 7-brane may require a different number of 5-branes ending on it than originally, which is due to the Hanany-Witten effect \cite{Hanany:1996ie}. 

Equivalently, the transition described above can be phrased as follows. First of all, let us define the intersection number between two 7-branes $j$ and $i$ as follows
\beq
\langle \ell_j,\ell_i \rangle = \det \left(\begin{array}{cc} p_j & q_j \\ p_i & q_i \end{array}\right)
\, .
\label{intersection_sides}
\eeq

The charge vectors satisfy the following relation
\beq
\sum_s N_i \ell_i =0 \, ,
\label{sum_charges}
\eeq
where $N_i$ is the number of 5-branes terminating on 7-brane $i$. Equation \eref{sum_charges} is simply the equilibrium condition for the $(p,q)$-web.

Crossing brane $j$ corresponds to changing the charge vectors as follows
\beq
\begin{array}{cll}
& \ell_j'  =  -\ell_j&  \\[.2cm]
i \neq j: \ \  & \ell_i' = \ell_i + \langle \ell_j,\ell_i \rangle \eta_j \ \ \ \ & \mbox{for } \langle \ell_j,\ell_i \rangle > 0 \\[.2cm]
& \ell_i' = \ell_i & \mbox{otherwise}
\end{array}
\label{mutation_polytope_1}
\eeq
In order to satisfy \eref{sum_charges} after the mutation, $N_j$ must transform according to
\beq
N_j' = \sum_{i \in L_+} N_i \langle \ell_j,\ell_i \rangle - N_{j}
\, ,
\label{mutation_polytope_2}
\eeq
where $L_{+}$ is the set of 7-branes with $\langle \ell_j,\ell_i \rangle > 0$. It is straightforward to show that \eref{mutation_polytope_1} is equivalent to \eref{monodromy}. There is an equivalent transformation in which, for a given crossed 7-brane $j$, the roles of the $\ell_i$'s with $\langle \ell_j,\ell_i \rangle >0$ and $\langle \ell_j,\ell_i \rangle < 0$ are exchanged in \eref{mutation_polytope_1}. More precisely, this corresponds to replacing $\langle \ell_j,\ell_i \rangle$ by $-\langle \ell_j,\ell_i \rangle$ everywhere in \eqref{mutation_polytope_1} and to moving the branch cut around the web in the opposite direction, i.e. using the inverse of the monodromy matrix.

\subsection{Geometric engineering and GTPs}

\begin{figure}[h!]
\centering
\includegraphics[height=4cm]{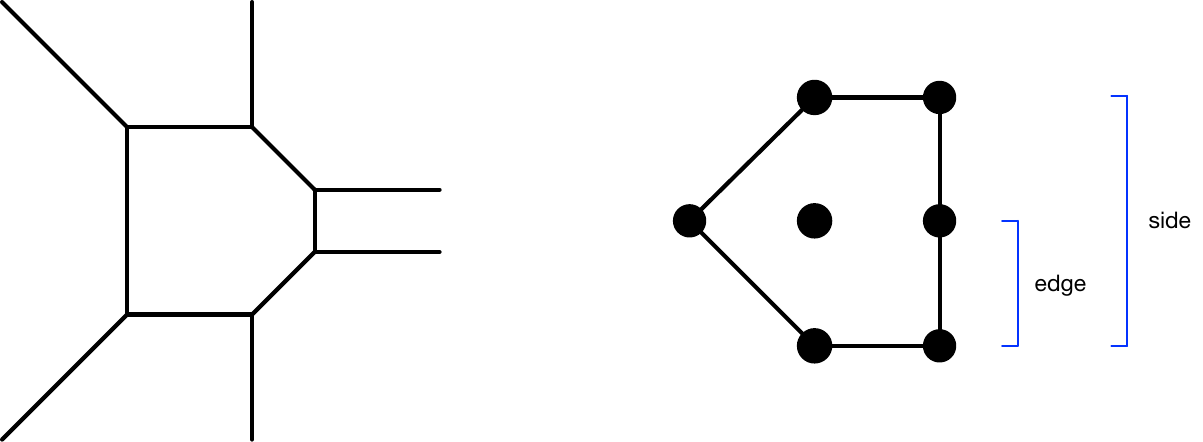}
\caption{A brane web and the corresponding toric diagram. The figure illustrates the concepts of edge and side of the toric diagram.}
\label{web_toric}
\end{figure}

Focusing on webs in which every 5-brane terminates on a different 7-brane, we can neglect the 7-branes and regard the external legs as semi-infinite segments. Such $(p,q)$-webs are related, via graph dualization, to toric diagrams of toric CY 3-folds.\footnote{More precisely, graph dualization connects a triangulation of the toric diagram to a $(p,q)$-web. Different triangulations map to different resolutions of the CY$_3$ and to different points on the extended Coulomb branch of the $5d$ theory.} Indeed, the same $5d$ theory can alternatively be obtained as the low energy limit of M-theory on $\mathbb{R}^{1,4}$ times the local CY 3-fold associated to the toric diagram \cite{Aharony:1997bh,Leung:1997tw}. From this perspective, the extended Coulomb branch corresponds to the K\"ahler cone of the CY 3-fold (see e.g. \cite{Closset:2020scj} for a detailed account). The interesting relations do not stop there, since the BPS spectrum of the $5d$ theory is encoded in a quiver theory that is precisely the one defined by the brane tiling corresponding to the toric diagram, i.e. the one for the $4d$ $\mathcal{N}=1$ gauge theory on the worldvolume of D3-branes probing the CY$_3$ \cite{Closset:2019juk}. \fref{web_toric} shows a $(p,q)$-web and the related toric diagram. 

While discussing the boundaries of toric diagrams, it is convenient to introduce the concepts of {\it sides} and {\it edges}. We refer to a side as the line connecting two consecutive corners of the toric diagram. Within a given side, an edge is a segment between two consecutive points in the toric diagram. The difference between the two types of objects becomes relevant for sides consisting of more than one edge.  Edges of the toric diagram are in one-to-one correspondence with legs of the dual $(p,q)$-web. Whenever a side of a toric diagram contains multiple edges, the $(p,q)$-web has the same number of parallel legs. \fref{web_toric} illustrates these ideas in an example. 

\begin{figure}[h!]
\centering
\includegraphics[height=5cm]{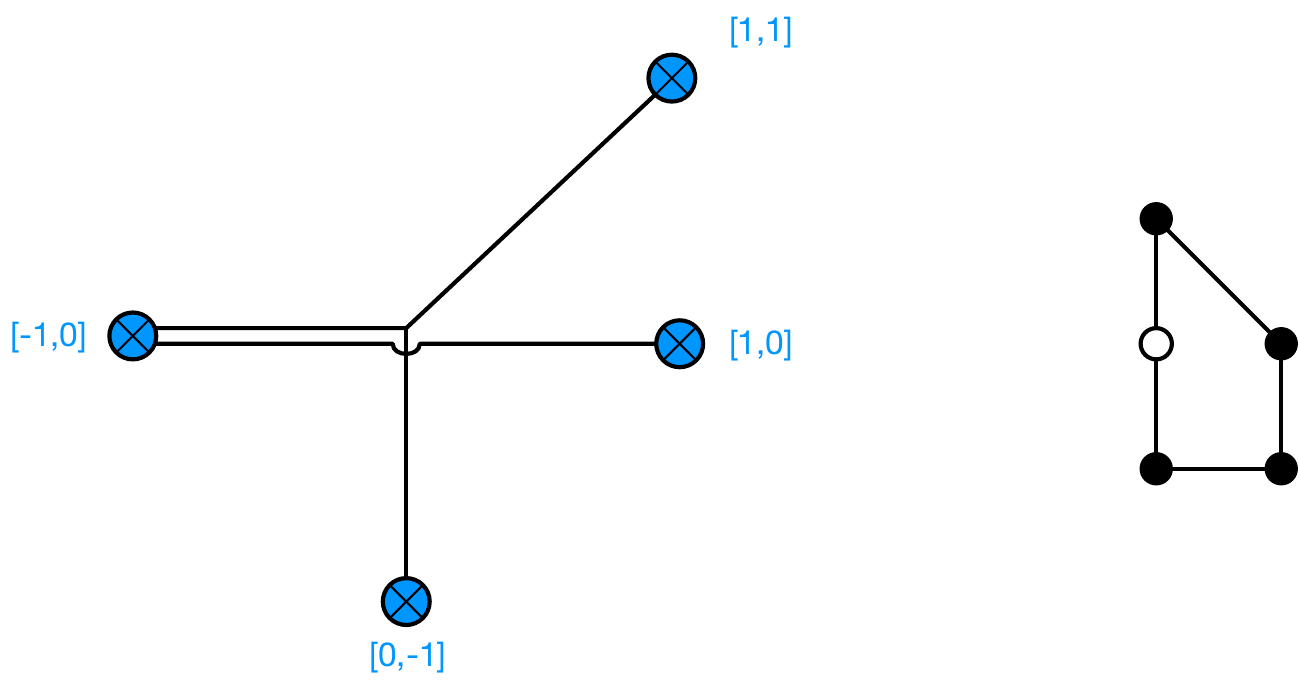}
\caption{A brane web in which multiple 5-branes terminate on the same 7-brane and the corresponding GTP. }
\label{GTP_web}
\end{figure}

In order to accommodate for multiplicities larger than 1, {\it generalized toric polygons} (GTPs) where proposed in \cite{Benini:2009gi} as generalizations of toric diagrams that encode such brane configurations. 7-branes are represented on the boundary of a GTP as follows. If two parallel legs of the $(p,q)$-web terminate on the same 7-brane, the dot that separates the corresponding edges in the GTP is colored white. Similarly, $n$ consecutive edges on a given side of the GTP separated by $n-1$ white dots represent $n$ parallel legs of the web terminating on a single 7-brane.\footnote{Similarly, white dots in the interior of a GTP capture the details of the 5-branes in the interior of the web, including the possible Coulomb branch directions. We refer the reader to  \cite{Benini:2009gi} as well as \cite{vanBeest:2020kou,VanBeest:2020kxw} for details.} \fref{GTP_web} shows a simple example.

\subsection{Constructing twin quivers}\label{constructingtwins}

\label{section_constructing_twin_quivers}

{\it Twin quivers} were introduced in \cite{Franco:2023flw} and provide a powerful new tool for the study of webs of 5- and 7-branes (equivalently GTPs) and the corresponding $5d$ theories.\footnote{As explained in \cite{Franco:2023flw}, the applications of twin quivers go well beyond such brane webs and $5d$ theories.} Twin quivers have a node for every 7-brane, whose rank is given by the number of 5-branes ending on it. A salient feature of twin quivers is that Hanany-Witten-type transitions on brane webs translate into quiver mutations, i.e. Seiberg duality.\footnote{We refer the reader to \cite{Feng:2002kk,Franco:2017lpa} to detailed discussions of Seiberg duality for quivers, including the transformation of the superpotential. In this paper we will formally apply the mutation rule even on $U(1)$ nodes of the quivers.} On a related note, they give an alternative perspective on the generalized $s$-rule that identifies supersymmetric brane configurations. In this section, we quickly review a method for constructing the twin quiver associated to general webs of 5-branes and 7-branes, equivalently to general GTPs, which was introduced in \cite{Franco:2023flw}. Twin quivers heavily rely on various objects and operations that are standard in the study of brane tilings and, more generally, bipartite field theories BFTs \cite{Franco:2012mm,Franco:2012wv,Franco:2013pg,Franco:2014nca,Franco:2018vqd}. They include zig-zag paths, untwisting and the connection between zig-zag paths and geometry. We refer the reader to \cite{Franco:2023flw} for a review of these ideas focused on applications to twin quivers, additional details and alternative algorithms for determining twin quivers. 

Let us consider a general brane web and denote $i$ the set of 5-brane legs with a given orientation $(p_{i},q_{i})$ and indicate the number of such legs as $N_{i}$.\footnote{In \cite{Franco:2023flw}, zig-zag paths and the corresponding nodes in twin quivers were indicated with tilded indices, to distinguish them from faces and quiver nodes in the {\it original} theories. Here, original refers to the theories defined by the brane tilings associated to the toric diagram under consideration, which are connected to the twin theories by untwisting. Since this paper is primarily devoted to twin quivers, we will omit the tildes to simplify the notation. In the few occasions in which we present the original theories, such as \fref{tilings_F0}, we expect that the meaning of indices will be clear from the context.} The configuration also contains $J_{i}\leq N_{i}$ $(p_{i},q_{i})$ 7-branes (denote each such leg by $\vec{\ell}^{(i)}_{A}=(p_{i},q_{i})$, with $A=1,\ldots, J_{i}$) on which these 5-branes can end. The multiplicities, i.e. the numbers of 5-branes terminating on each of the 7-branes, are $k^{(i)}_{A}=\{k^{(i)}_1,\ldots,k^{(i)}_{J_{i}}\}$. We can alternatively discuss such a web in GTP language, where it maps to a polytope with a general array of black and white dots on its boundary. Each side $i$ contain $N_{i}$ edges, which are partitioned into subsets containing $\{k^{(i)}_1,\ldots,k^{(i)}_{J_{i}}\}$ edges, such that the edges in each subset are separated by white dots and different subsets are separated by black dots.

\begin{figure}[h!]
\centering
\includegraphics[height=6cm]{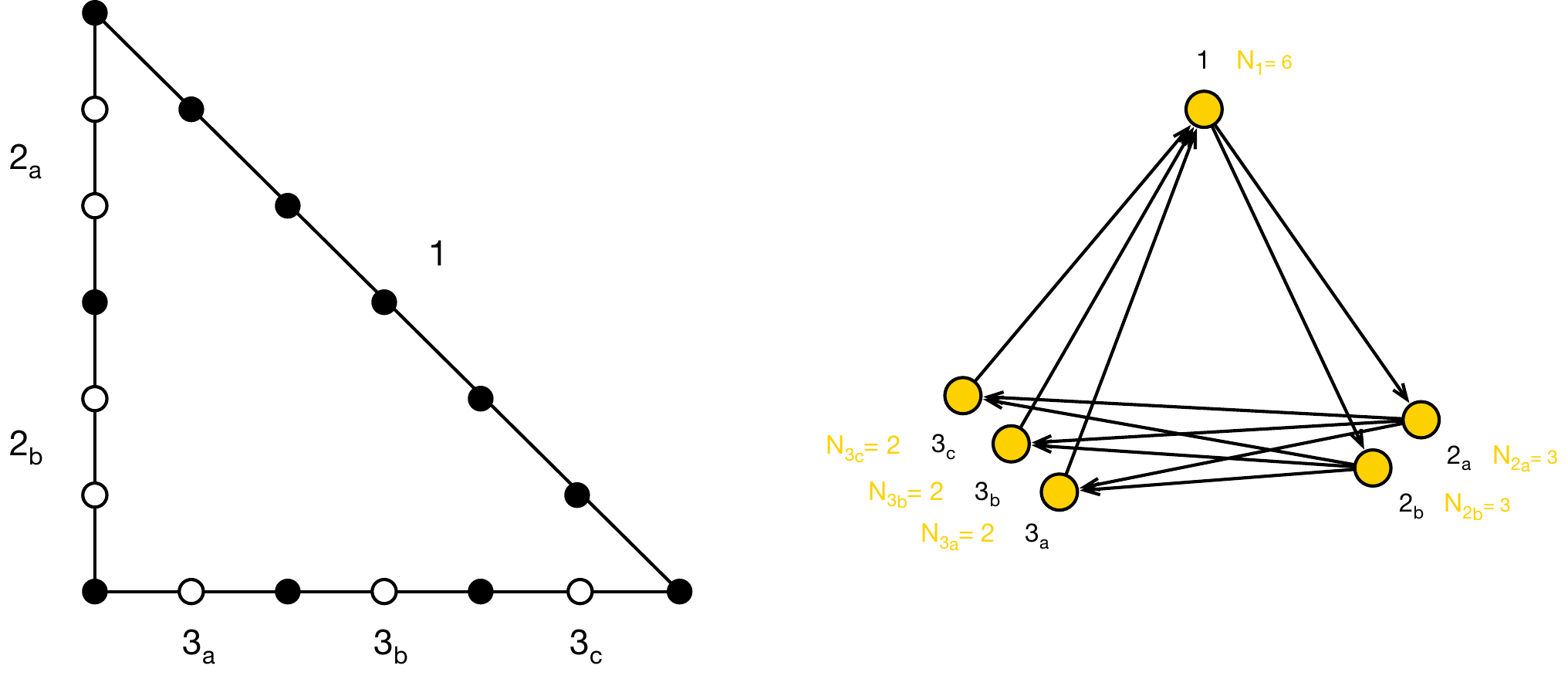}
\caption{A GTP, for which we only show the boundary, and the corresponding twin quiver.} 
\label{example_twin_quiver}
\end{figure}

The associated twin quiver is constructed as follows:

\begin{enumerate}
\item Regarding the polytope under consideration as a toric diagram (namely with all black dots), construct the corresponding brane tiling. 
\item Generate the BFT for $\tilde{Q}_{\mbox{T}}$ by untwisting.\footnote{By this, we mean the corresponding bipartite graph on a Riemann surface.} This quiver has one node for every zig-zag path of the original tiling/edge of the original toric diagram. 
\item Finally, the $N_{i}$ nodes associated to every side with $N_{i}$ edges in the original toric diagram are merged into $J_{i}$ nodes of ranks $k^{(i)}_1,\ldots,k^{(i)}_{J_{i}}$. 
\end{enumerate}
While we have primarily focused on the quiver diagram, the procedure outlined above can be refined to also produce the superpotential of the resulting twin quiver theory.\footnote{This algorithm will be reported in future work.} \fref{example_twin_quiver} presents an example illustrating basic features of this construction.

As anticipated above, the twin quiver constructed in this way has a $U(k^{(i)}_A)$ gauge group for each external leg with charge vector $\vec{\ell}^{(i)}_{A}$ and $N_{A,B}^{i,j}=\langle \vec{\ell}^{(i)}_{A},\vec{\ell}^{(j)}_{B} \rangle$ bifundamentals from node $U(k^{(i)}_A)$ to node $U(k^{(j)}_B)$ if $N_{A,B}^{i,j}>0$ and in the opposite direction if $N_{A,B}^{i,j}<0$.

\section{Twin quivers for different toric phases} 

\label{section_twin_quivers_for_different_toric_phases} 

In the previous section, we reviewed the construction of the twin quiver associated to a GTP/brane web. The construction involves the determination of a brane tiling/quiver theory $Q$ for a standard toric diagram, which is followed by untwisting and certain identifications in the case of GTPs containing white dots. Generically, a given toric diagram is associated to multiple brane tilings. These different quiver theories are known as toric phases and are connected to each other by mutations on toric nodes. It is then natural to ask how the non-uniqueness of $Q$ reflects on the twin quivers $\tilde{Q}$. These issue was first noticed and addressed in \cite{Franco:2023flw}, to where we refer the reader for further details.

\begin{figure}[h!]
\centering
\includegraphics[height=5cm]{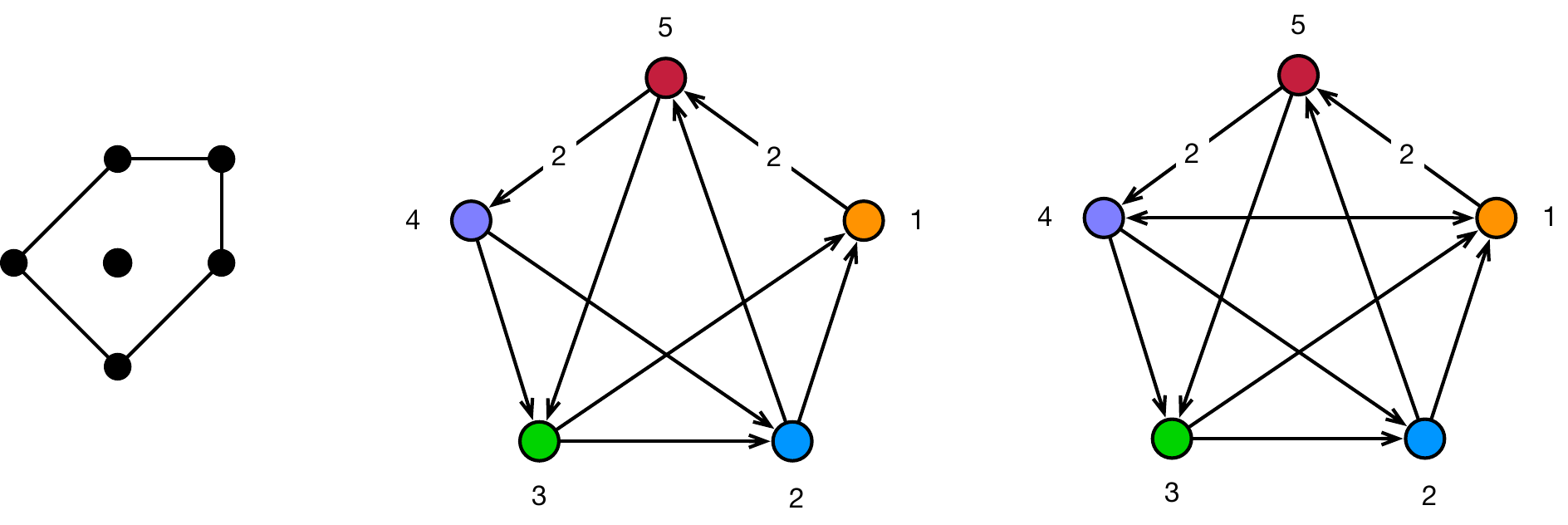}
\caption{The $dP_2$ geometry has two toric phases, which gives rise to the two twin quivers shown in this figure upon untwisting.}
\label{phases_twin_quivers_dP2}
\end{figure}

In \cite{Franco:2023flw} it was shown that the twin quivers for a given GTP constructed using different toric phases of $Q$ differ by bidirectional arrows.\footnote{It is also possible to understand the superpotentials of these different twin quivers, but we leave a more detailed discussion for future work.} This phenomenon can be understood in general using the construction of twin quivers of Section \sref{sect:twin_quivers}. After replacing the GTP by a toric diagram (if necessary), let us refer to the two toric phases associated to it as $Q_{\rm{T},(a)}$ and $Q_{\rm{T},(b)}$. The (sequence of) Seiberg duality transformation(s) connecting $Q_{\rm{T},(a)}$ and $Q_{\rm{T}(b)}$ corresponds to a reorganization of some of the zig-zag paths that preserves their homology \cite{Hanany:2005ss}. In this process, the intersections between zig-zag paths change, but they appear/disappear in pairs, with opposite signs. Since nodes in $\tilde{Q}_{\rm{T},(a)}$ and $\tilde{Q}_{\rm{T}(b)}$ correspond to zig-zags, they only differ by bidirectional arrows. This property is preserved if the GTP contains white dots and, correspondingly, nodes in the twin quiver are combined into higher rank ones to form $\tilde{Q}_{\rm{GTP},(a)}$ and $\tilde{Q}_{\rm{GTP}(b)}$. \fref{phases_twin_quivers_dP2} illustrate the non-uniqueness of the twin quivers in an explicit example, in which the GTP is simply the standard toric diagram for $dP_2$. This example will be revisited in Section \sref{dP2}.

It is important to emphasize that, as the example in \fref{phases_twin_quivers_dP2} illustrates, the non-uniqueness of the twin quivers is a generic phenomenon, which is present even for ordinary toric diagrams/web configurations. In other words, it is unrelated to whether the brane web have multiple legs terminating on the same 7-brane or not. This is also reflected in the fact that the $N_{A,B}^{i,j}$ introduced above are only sensitive to the antisymmetric part of the connection matrix, and thus insensitive to bidirectional arrows in the quiver.

In this paper we will investigate the physical significance of the multiple twin quivers, i.e. what properties or the corresponding brane webs they capture. We will see that these capture different regions of the extended Coulomb branch of the $5d$ theory.

\section{Conifold/$\mathbb{Z}_N$}\label{sect:conifold/Zn}

\begin{figure}[h!]
\centering
\includegraphics[height=6cm]{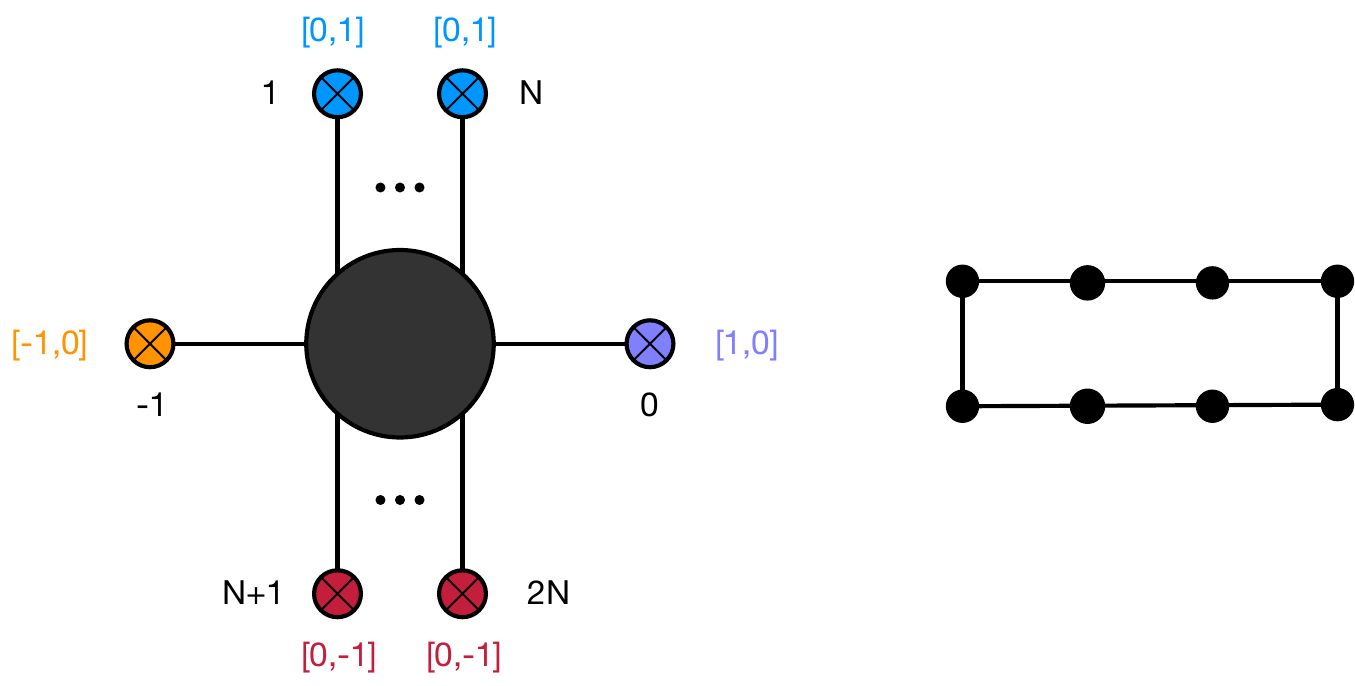}
\caption{Brane web for $N^2$ hypermultiplets. On the right, the dual toric diagram is the one for a non-chiral $\mathbb{Z}_N$ orbifold of the conifold (here shown for $N=3$).}
\label{web_ZN_conifold}
\end{figure}

Let us consider the brane web shown in \fref{web_ZN_conifold}, which describes $N^2$ free hypermultiplets in $5d$. This web can also be regarded as the $M=1$ case of the $+_{N,M}$ family of theories introduced in \cite{Bergman:2018hin}. The corresponding GTP, is the ordinary toric diagram for a non-chiral $\mathbb{Z}_N$ orbifold of the conifold.

\begin{figure}[h!]
\centering
\includegraphics[height=6cm]{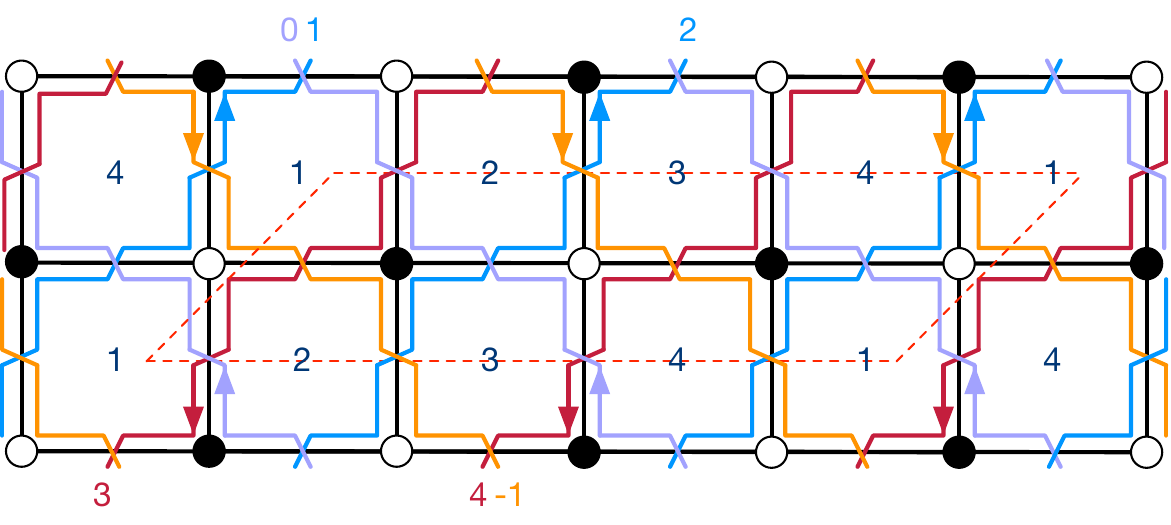}
\caption{Brane tiling for the non-chiral $\mathbb{Z}_N$ orbifold of the conifold under consideration for the case $N=2$. We show the zig-zag paths associated to the legs of the web in \fref{web_ZN_conifold}. Dashed red lines indicate the boundary of the unit cell.}
\label{tiling_ZN_conifold}
\end{figure}

It is straightforward to construct the corresponding brane tiling, which is shown in \fref{tiling_ZN_conifold}.\footnote{The winding numbers of the zig-zag paths correspond to the $(p,q)$ charges of the corresponding legs in the web. Other equivalent choices of the unit cell result in $(p,q)$ charges that differ from the ones in \fref{web_ZN_conifold} by an $SL(2,\mathbb{Z})$ transformation.} It is obtained by appending $N$ copies of the unit cell describing the conifold theory. The orbifold action under consideration translates into the details of how these copies are combined (see e.g. \cite{Hanany:2005ve,Franco:2005rj,Davey:2010px}).

\begin{figure}[h!]
\centering
\includegraphics[height=5.5cm]{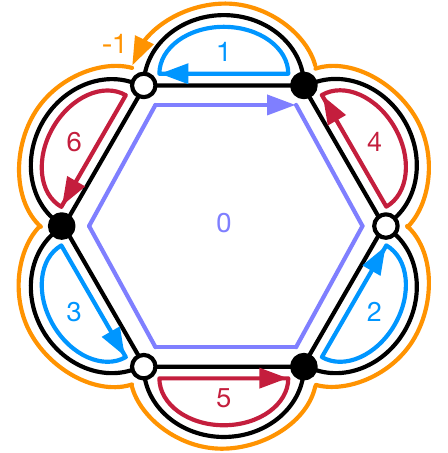}
\caption{Bipartite graph on a sphere with $2N+2$ punctures describing the twin quiver theory. The polygon at the center is a $2N$-gon. Here we show the case of $N=3$.}
\label{conifoldmodZNuntwisteddimer}
\end{figure}

The twin theory is obtained by untwisting the brane tiling. As expected from the corresponding web and toric diagrams, it corresponds to a bipartite graph on a sphere with $2N+2$ punctures, each of which can be identified with a leg of the web. \fref{conifoldmodZNuntwisteddimer} shows the untwisted graph for $N=3$.

\begin{figure}[h!]
\centering
\includegraphics[height=5.5cm]{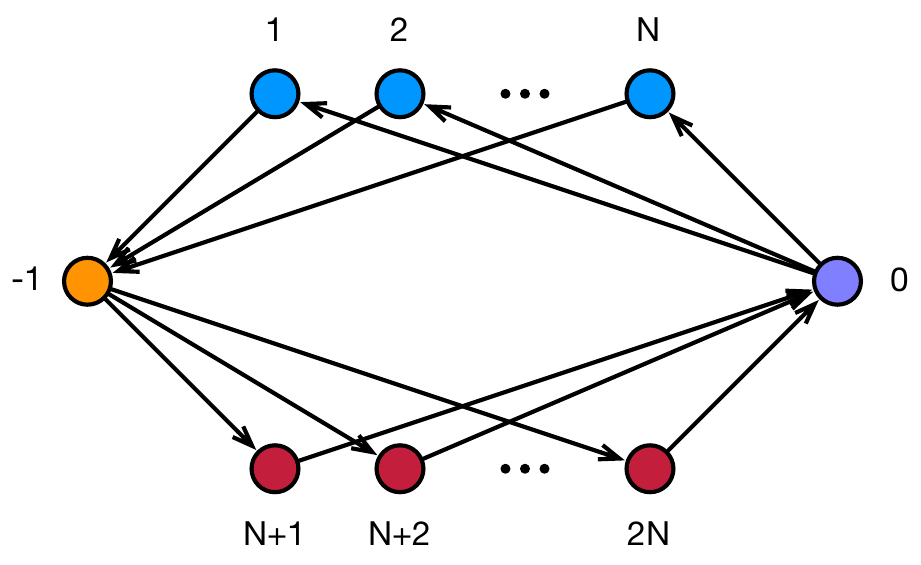}
\caption{Twin quiver for a non-chiral $\mathbb{Z}_N$ orbifold of the conifold, which corresponds to $N^2$ free hypermultiplets in $5d$.}
\label{N_hypers_quiver}
\end{figure}

The twin quiver, shown in \fref{N_hypers_quiver}, has a node for every 7-brane in the configuration. In order simplify figures, throughout the paper the rank of quiver nodes will be 1 unless explicitly indicated. The superpotential is quartic and can be easily read from \fref{conifoldmodZNuntwisteddimer}. Below we write it for some explicit examples.

\subsection{Conifold/$\mathbb{Z}_2$}\label{conifold/Z2}

\begin{figure}[h!]
\centering
\includegraphics[height=5.5cm]{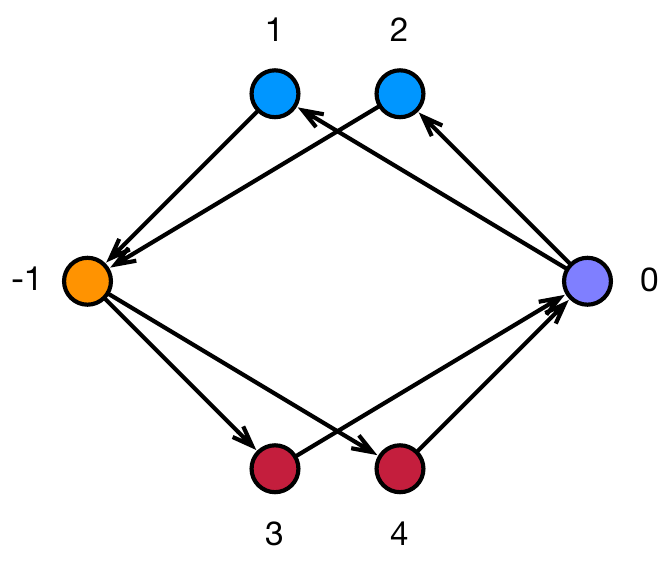}
\caption{Twin quiver for a non-chiral $\mathbb{Z}_2$ orbifold of the conifold, which corresponds to $4$ free hypermultiplets in $5d$.}
\label{4_hypers_quiver}
\end{figure}

For concreteness, let us consider the case of $N=2$. The twin quiver is shown in \fref{4_hypers_quiver}.

The superpotential for the twin quiver can be read from the untwisted dimer, analogous to \fref{conifoldmodZNuntwisteddimer}, and is
\begin{eqnarray}
    W &=&X_{1,-1}X_{-1,4}X_{4,0}X_{0,1}+X_{3,0}X_{0,2}X_{2,-1}X_{-1,3} \\ \nonumber
	&-&X_{0,1}X_{1,-1}X_{-1,3}X_{3,0}-X_{-1,4}X_{4,0}X_{0,2}X_{2,-1}\,.
\end{eqnarray}

From \fref{4_hypers_quiver}, it is clear that there are two qualitatively different possible mutations: either mutating one of the $-1,0$ nodes, or mutating one of the $1,\ldots, 4$ nodes. Let us consider each case in further detail.

\bigskip

\begin{figure}[h!]
\centering
\includegraphics[height=6cm]{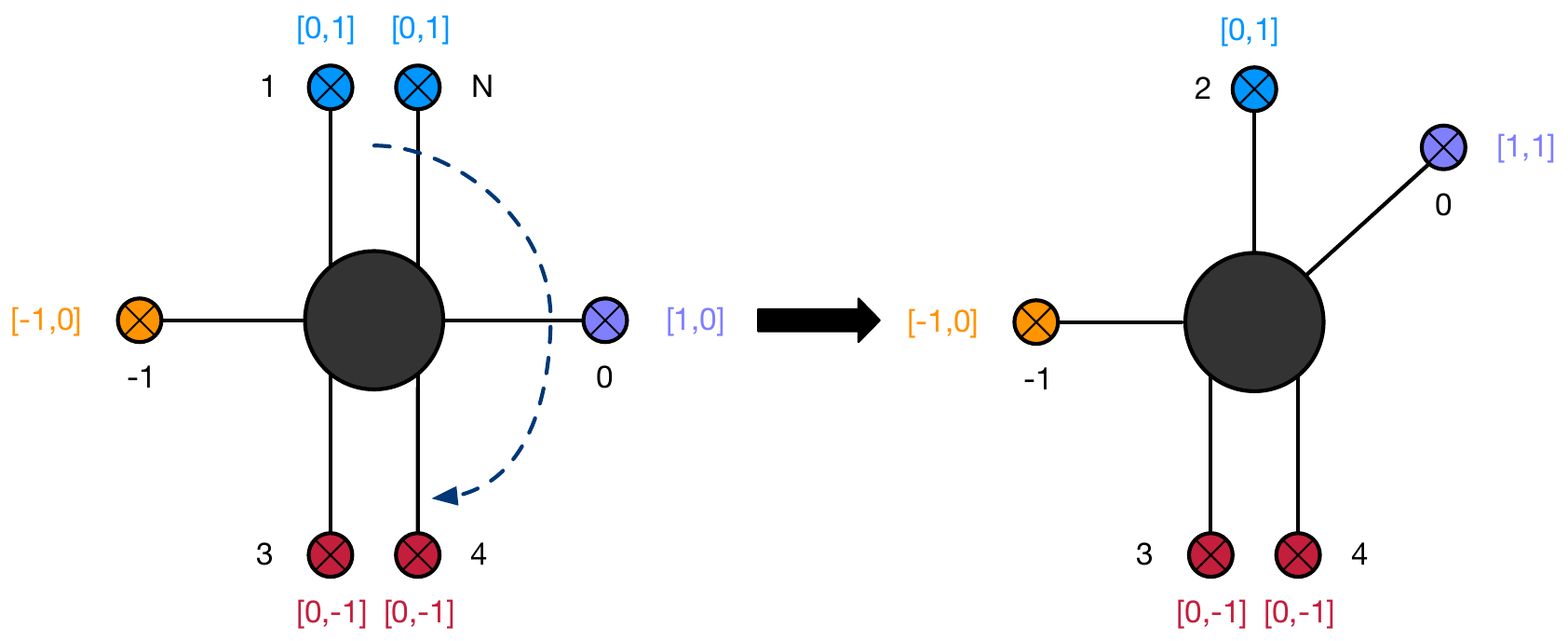}
\caption{Brane crossing connecting \fref{web_ZN_conifold} to \fref{3_hypers}.}
\label{web_crossing_Z2_conifold}
\end{figure}

\begin{figure}[h!]
\centering
\includegraphics[height=6cm]{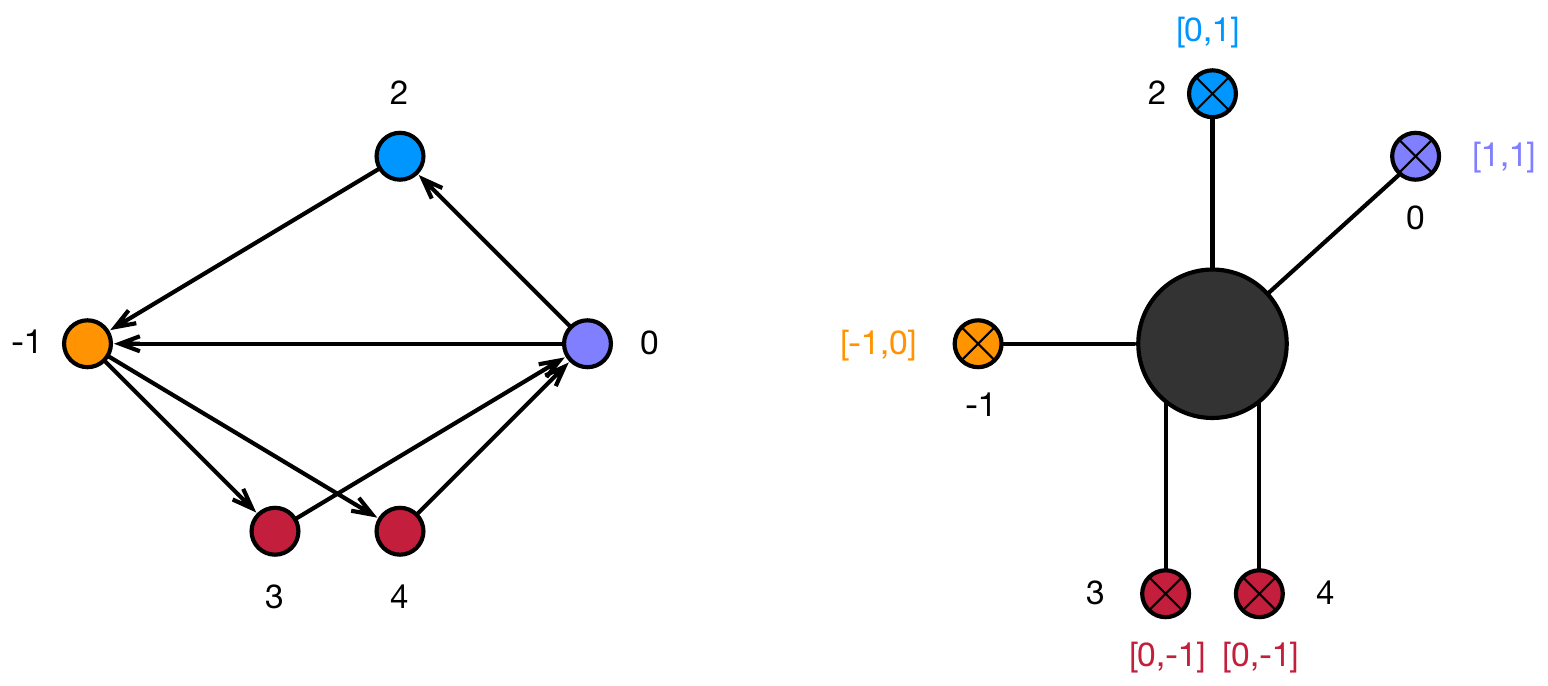}
\caption{Quiver and web after mutating node 1 in \fref{4_hypers_quiver}.}
\label{3_hypers}
\end{figure}

\paragraph{Mutating $1$.} Since node 1 has $N_f=N_c=1$, it disappears upon mutation. As a consequence, there are no magnetic quarks. There is only one new field in the quiver, the meson $X_{0,-1}$, which, in terms of fields in the theory before the mutation, corresponds to the composition $X_{0,1}X_{1,-1}$. The resulting quiver is shown on the left panel of \fref{3_hypers}. It indeed corresponds to the brane web on the right of \fref{3_hypers}, which was obtained from the one in \fref{web_ZN_conifold} by the brane crossing shown in \fref{web_crossing_Z2_conifold}. We will perform similar brane crossings in the other examples considered in this paper.
    
The superpotential follows from the mutation rules and is 
\begin{eqnarray}
    W&=&X_{0,-1}X_{-1,4}X_{4,0}+X_{3,0}X_{0,2}X_{2,-1}X_{-1,3} \\ \nonumber
    &-&X_{0,-1}X_{-1,3}X_{3,0}-X_{-1,4}X_{4,0}X_{0,2}X_{2,-1}\,.
\end{eqnarray}

\begin{figure}[h!]
\centering
\includegraphics[height=6cm]{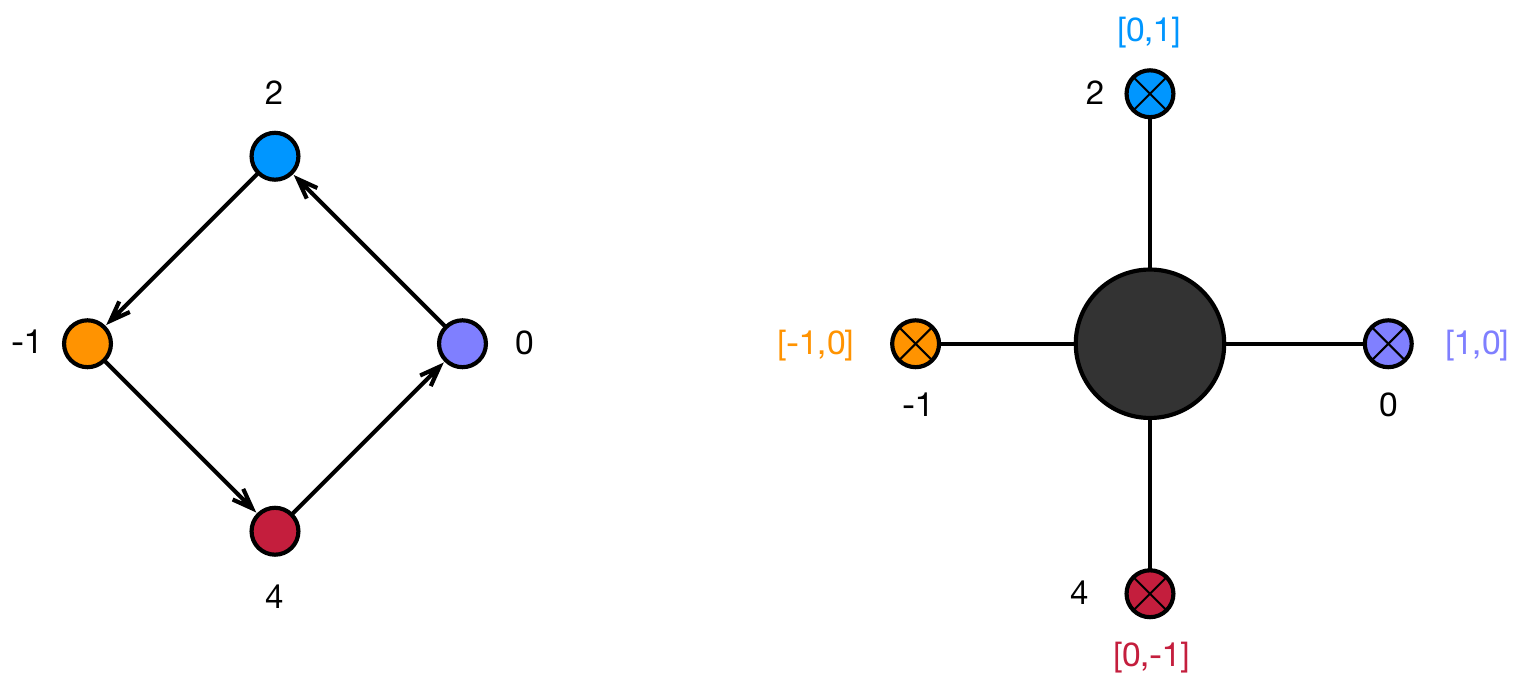}
\caption{Quiver and web after mutating nodes 1 and 3 in \fref{4_hypers_quiver}.}
\label{1_hyper}
\end{figure}

It is interesting to perform one further mutation on node 3. Once again, $N_f=N_c=1$ for the mutated node, so it disappears upon mutation. The new meson is $Y_{-1,0}=X_{-1,3}X_{3,0}$. The superpotential is given by 

\begin{eqnarray}
    W&=&X_{0,-1}X_{-1,4}X_{4,0}+X_{0,2}X_{2,-1}Y_{-1,0}\\ \nonumber
    &-&X_{0,-1}Y_{-1,0}-X_{-1,4}X_{4,0}X_{0,2}X_{2,-1}\,.
\end{eqnarray}
Integrating out the massive fields $X_{0,-1}$ and $Y_{-1,0}$, we obtain the quiver in \fref{1_hyper} and the superpotential vanishes.\footnote{More precisely, the untwisted dimer (as the original one) has two vertices of order 4. Those two vertices would correspond to two quartic terms involving the same fields, but in different order. Since the twin quiver under consideration is Abelian, the ordering does not matter and the two terms cancel each other. The structure of the superpotential can also be determined using the construction in Section \sref{constructingtwins}.} As expected, this twin quiver is the one associated to crossing upwards one of the red 7-branes in the web on the left panel of \fref{3_hypers}. Finally, we can directly obtain this twin quiver starting from the 
the dimer in \fref{conifoldmodZNuntwisteddimer} for the $N=1$ case.

\bigskip

\paragraph{Mutating $-1$.} 
This node has $N_f=2N_c =1$. As a result, the rank of node $-1$ is still 1 after the mutation.  The twin quiver and web after the mutation are shown in \fref{T2} below. They correspond to the so-called $T_2$ theory. The superpotential is 
 \begin{eqnarray}
    \label{Wforconifold/Z2mutated}
W & = & X_{1,4}X_{4,0}X_{0,1}+X_{3,0}X_{0,2}X_{2,3} 
     -  X_{0,1}X_{1,3}X_{3,0} - X_{4,0}X_{0,2}X_{2,4} \\ \nonumber
    &+ & X_{-1,1}X_{1,3}X_{3,-1} - X_{-1,1}X_{1,4}X_{4,-1} - X_{-1,2}X_{2,3}X_{3,-1}+X_{-1,2}X_{2,4}X_{4,-1}\,.
\end{eqnarray}

\begin{figure}[h!]
\centering
\includegraphics[height=6cm]{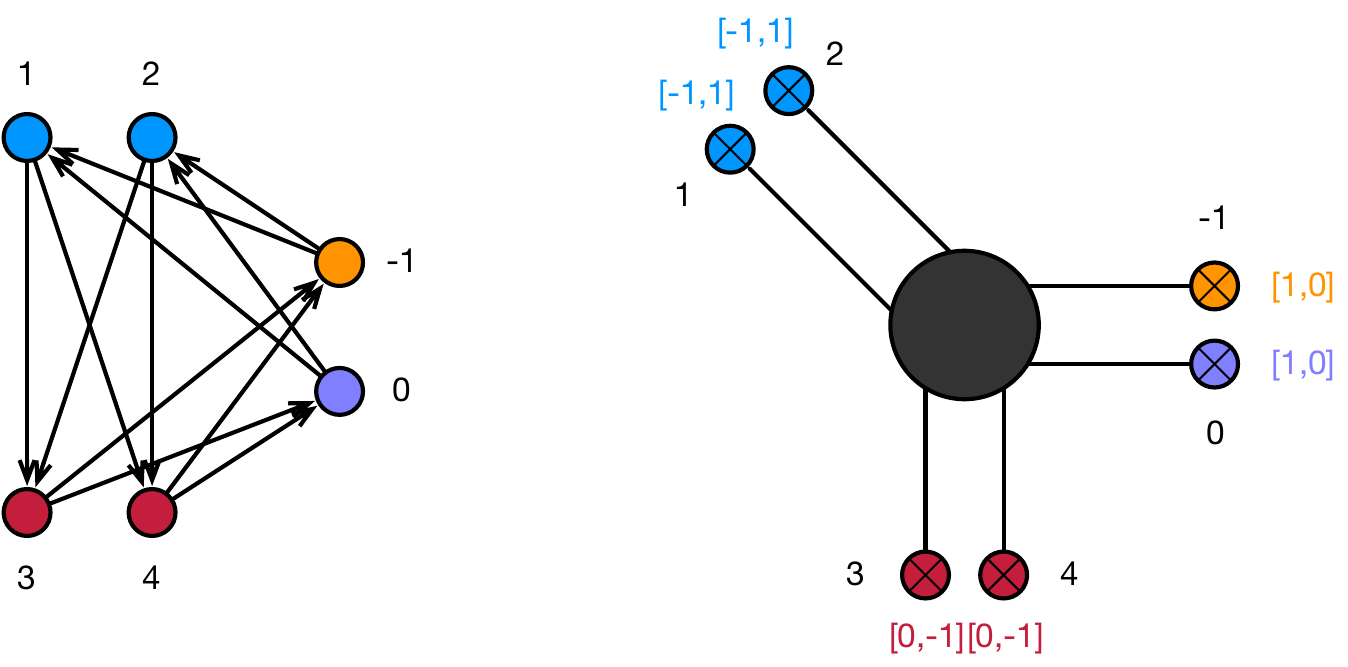}
\caption{Quiver and web after mutating node $-1$ in \fref{4_hypers_quiver}.}
\label{T2}
\end{figure}

\section{Multiple toric phases, a first encounter}

\label{section_multiple_toric_phases}

The alert reader might notice that the starting point of our analysis in Section \sref{sect:conifold/Zn} was rather arbitrary. In particular, the $\mathbb{Z}_N$ orbifold of the conifold has multiple toric phases for $N>1$, and the brane tiling in \fref{tiling_ZN_conifold} is just one of them. The toric phase described by \fref{tiling_ZN_conifold} is ``minimal", in the sense of having the minimum number of chiral fields and being obtained by combining copies of the conifold brane tiling. However, it is natural to ask how the results of the previous section are affected by starting from a different phase.

\begin{figure}[h!]
\centering
\includegraphics[height=6cm]{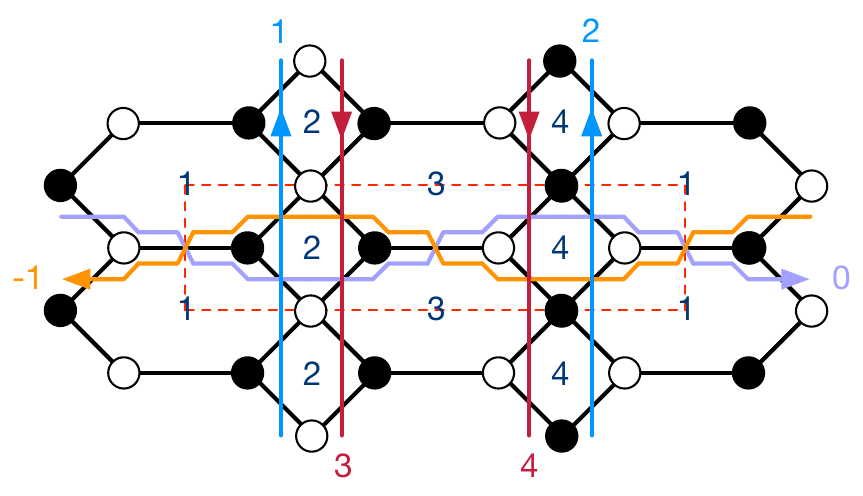}
\caption{Brane tiling for another toric phase of the non-chiral $\mathbb{Z}_2$ orbifold of the conifold under consideration. We show the zig-zag paths associated to the legs of the web in \fref{web_ZN_conifold}.}
\label{tiling_Z2_conifold_phase_2}
\end{figure}

\begin{figure}[h!]
\centering
\includegraphics[height=5cm]{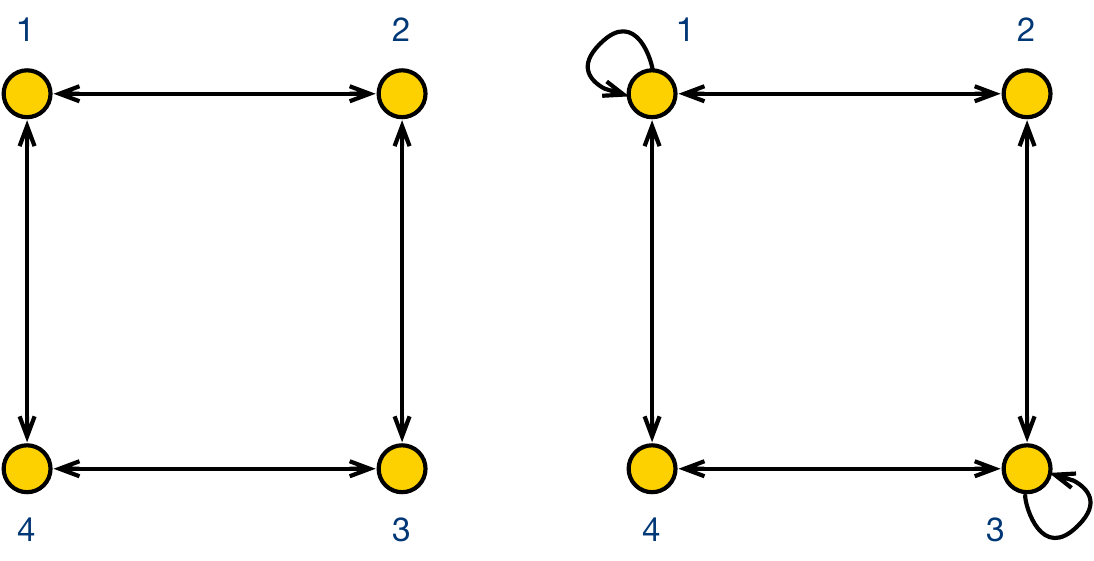}
\caption{Original quivers for the two toric phases of the non-chiral $\mathbb{Z}_2$ orbifold of the conifold under consideration. They are connected by Seiberg duality on either node 2 or 4. Mutating node 1 or 3 of the first quiver leads to equivalent results.} 
\label{quivers_Z2_conifold}
\end{figure}

For concreteness, let us continue with the case of $N=2$. This geometry has only two toric phases, up to relabeling of the nodes. One of them is the one we have previously considered, which is given by \fref{tiling_ZN_conifold}, while the new one is described by the brane tiling in \fref{tiling_Z2_conifold_phase_2}. \fref{quivers_Z2_conifold} shows the corresponding quiver

\begin{figure}[h!]
\centering
\includegraphics[height=6cm]{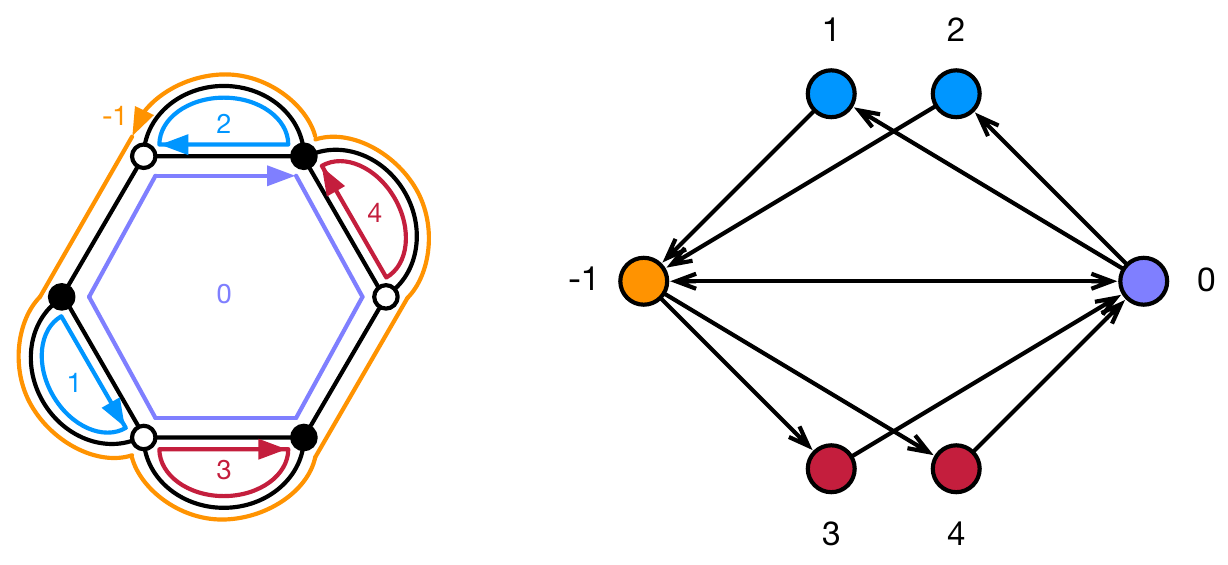}
\caption{Untwisted dimer and corresponding twin quiver for the second phase of the non-chiral $\mathbb{Z}_2$ orbifold of the conifold associated to 4 free hypermultiples in $5d$.}
\label{4_hypers_phase_2}
\end{figure}

Let us now repeat the analysis of Section \sref{conifold/Z2}, but using the toric phase in \fref{tiling_Z2_conifold_phase_2} as the starting point. Untwisting leads to the twin quiver defined by the bipartite graph in \fref{4_hypers_phase_2}. As expected, the graph lives on a sphere with 6 punctures. However, it differs from the $N=2$ case of the graph shown in \fref{conifoldmodZNuntwisteddimer}. On the right of the figure, we show the corresponding quiver. 

The difference between the twin quivers in Figures \ref{4_hypers_quiver} and \ref{4_hypers_phase_2} is the presence of bidirectional arrows.\footnote{The two theories, of course, also differ in the superpotentials. These can be read off the corresponding bipartite graphs.} This could have been in no other way, since the antisymmetric part of the adjacency  
matrix can be determined from the web in \fref{web_ZN_conifold} and must be the same. The superpotential for the twin quiver is
\begin{eqnarray}
    W & =& X_{-1,0}X_{0,1}X_{1,-1}+X_{-1,3}X_{3,0}X_{0,-1}+X_{-1,4}X_{4,0}X_{0,2}X_{2,-1} \\ \nonumber
        & - & X_{-1,3}X_{3,0}X_{0,1}X_{1,-1} - X_{-1,0}X_{0,2}X_{2,-1} - X_{-1,4}X_{4,0}X_{0,-1}\,.
\end{eqnarray}

Below, we study the mutations of the web. As before, we have two qualitatively different possibilities.

\bigskip

\begin{figure}[h!]
\centering
\includegraphics[height=6cm]{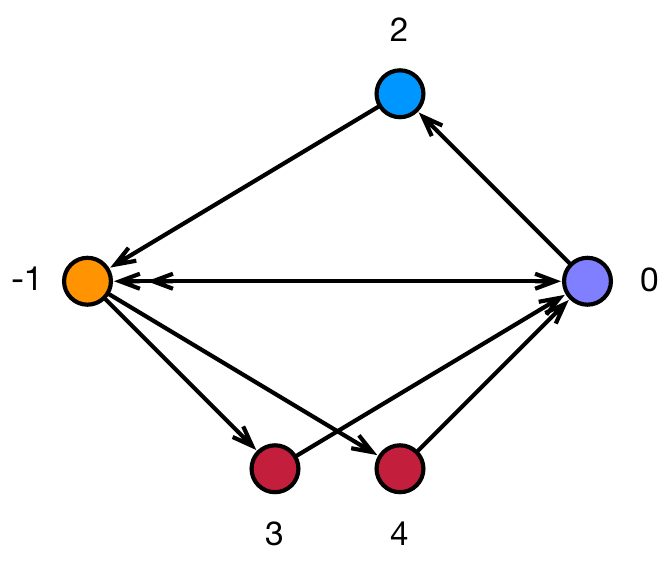}
\caption{Quiver after mutating node 1 in \fref{4_hypers_phase_2}. A pair of arrows connecting nodes 0 and -1 in opposite directions are massive and can be integrated out.}
\label{3_hypers_phase_2_before_integrating_out}
\end{figure}

\paragraph{Mutating $1$.} 
Let us mutate the twin quiver on any of the blue or red nodes in \fref{4_hypers_phase_2}. Without loss of generality, we can assume it is node 1. Since the extra bidirectional arrow is not connected to node 1, the mutation is qualitatively similar to the previous case. Once again, since $N_f=N_c=1$, node 1 disappears in the mutation. There is a new meson $Y_{0,-1}=X_{0,1}X_{1,-1}$ and the superpotential is given by
\begin{eqnarray}
    W & = & Y_{0,-1}X_{-1,0}+X_{-1,3}X_{3,0}X_{0,-1}+X_{-1,4}X_{4,0}X_{0,2}X_{2,-1}\\ \nonumber
    & - & X_{-1,3}X_{3,0}Y_{0,-1} - X_{-1,0}X_{0,2}X_{2,-1} - X_{-1,4}X_{4,0}X_{0,-1}\,.
\label{W_twin_conifold_Z2_phase2_mutation_on_1}
\end{eqnarray}
\fref{3_hypers_phase_2_before_integrating_out} shows the mutated twin quiver.

From the superpotential \eqref{W_twin_conifold_Z2_phase2_mutation_on_1}, we see that the fields $Y_{0,-1}$ and $X_{-1,0}$ in the bidirectional arrow are massive. Integrating them out, we remove the bidirectional arrow from the quiver and the superpotential becomes
\begin{eqnarray}
    W & = & X_{-1,3}X_{3,0}X_{0,-1}+X_{-1,4}X_{4,0}X_{0,2}X_{2,-1} \\ \nonumber
    & - & X_{-1,3}X_{3,0}X_{0,2}X_{2,-1} - X_{-1,4}X_{4,0}X_{0,-1}\,.
\end{eqnarray}
This is precisely the theory in \fref{3_hypers}, upon the relabeling of nodes $3\leftrightarrow 4$.

\paragraph{Mutating $-1$.} 

\begin{figure}[h!]
\centering
\includegraphics[height=6cm]{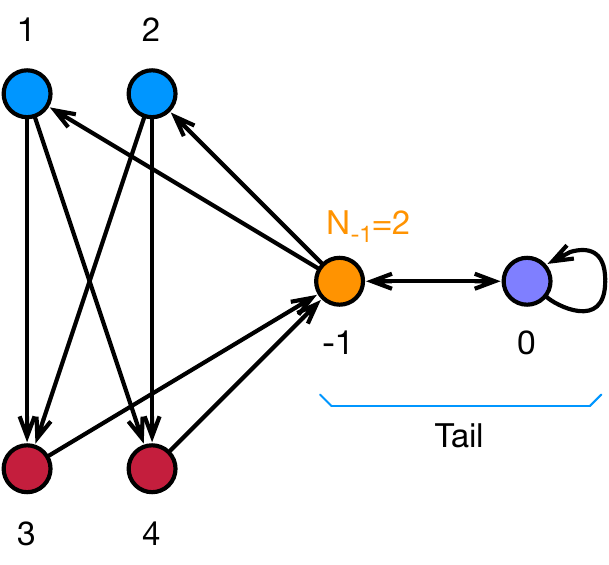}
\caption{Quiver obtained by mutating \fref{4_hypers_phase_2} on node $-1$. This example contains a quiver tail.}
\label{2_hypers_phase2_mutated}
\end{figure}

Things become more interesting when we mutate node $-1$ (or equivalently node 0), since the additional bidirectional arrow increases its number of flavors. Indeed, node $-1$ has $N_f=3N_c=3$ and it therefore becomes a $U(2)$ gauge group after mutation. The incoming fields $X_{A,-1}$ ($A=\{0,1,2\}$) combine with the outgoing ones $X_{-1,a}$ ($a=\{0,3,4\}$) 
to form mesons $M_{A,a}=X_{A,-1}X_{-1,a}$. Finally, we denote the magnetic quarks $Y_{-1,A}$ and $Y_{a,-1}$. The resulting superpotential is 
\begin{eqnarray}
W & = & X_{0,1}M_{1,0}+X_{3,0}M_{0,3}+X_{4,0}X_{0,2}M_{2,4} \\ \nonumber
    & - & X_{3,0}X_{0,1}M_{1,3} - X_{0,2}M_{2,0} - X_{4,0}M_{0,4} \\ \nonumber
   & + & Y_{-1,A}M_{A,a}Y_{a,-1}\,.
\end{eqnarray}
The fields $X_{0,1}$, $X_{0,2}$ ,$X_{3,0}$ , $X_{4,0}$, $M_{1,0}$, $M_{2,0}$, $M_{0,3}$ and $M_{4,0}$ are massive. Integrating them out, we obtain the quiver in 
\fref{2_hypers_phase2_mutated}, whose superpotential is
\begin{eqnarray}
W&=&Y_{0,-1}Y_{-1,0}M_{0,0}-Y_{-1,0}Y_{0,-1}Y_{-1,1}M_{1,3}Y_{3,-1}+Y_{-1,0}Y_{0,-1}Y_{-1,2}M_{2,4}Y_{4,-1} \nonumber \\ \nonumber && +M_{1,3}Y_{3,-1}Y_{-1,1}+M_{2,3}Y_{3,-1}Y_{-1,2}+M_{1,4}Y_{4,-1}Y_{-1,1}+M_{2,4}Y_{4,-1}Y_{-1,2}\,.\\ &&
\label{W_quiver_tail_1}
\end{eqnarray}

The twin quiver in \fref{2_hypers_phase2_mutated} exhibits an interesting structure, which we will denote {\it quiver tail}. Node $0$ contains an adjoint and is connected to node $-1$ by a bidirectional arrow. Moreover, the ranks of the nodes decrease along the tail.

\begin{figure}[h!]
\centering
\includegraphics[height=6cm]{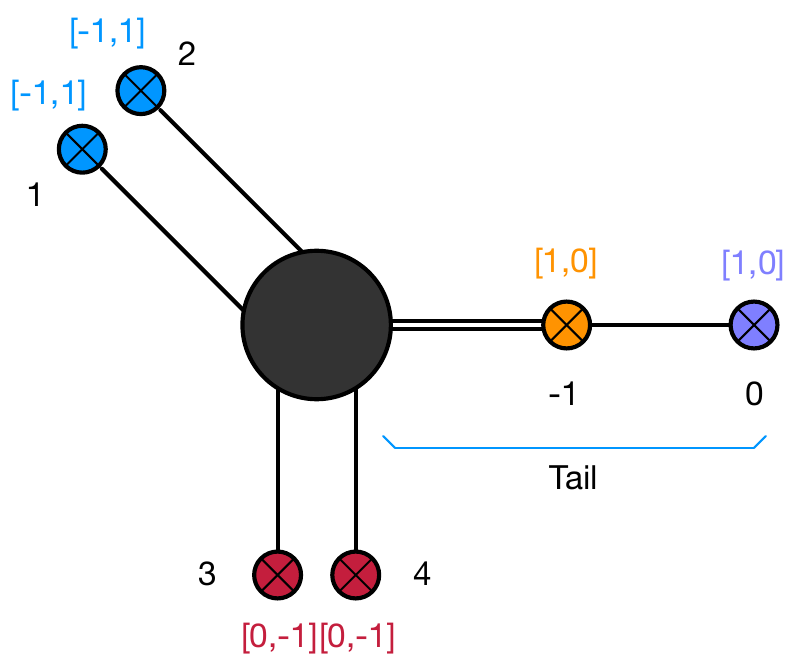}
\caption{Brane web associated to the twin quiver with a tail in \fref{2_hypers_phase2_mutated}.}
\label{Web_2_Hypers_Phase2_Mutated}
\end{figure}

What does the tail in the twin quiver represent for the corresponding brane web? We claim that its natural interpretation is given by \fref{Web_2_Hypers_Phase2_Mutated}. This web is a particular mass deformation of the one in \fref{T2}, in which masses are tuned such that two 7-branes are aligned so the corresponding 5-branes can split into segments while respecting the $s$-rule. The existence of the quiver tail can be traced to the fact that we started from a different toric phase for the underlying geometry. This suggests that different toric phases of the original theory allow to capture the roots of the Higgs branch in the extended Coulomb branch of the associated $5d$ theory. In what follows, we will investigate this proposal in further detail.

\subsection{Tail decoupling, GTPs and node merging} \label{TailDecouplingAndNodeMerging}

Motivated by the previous results, let us investigate how twin quivers capture other transformations of the corresponding webs. Let us strip off the tail in \fref{Web_2_Hypers_Phase2_Mutated}. To do so, we turn on a non-zero VEV for the adjoint in \fref{2_hypers_phase2_mutated}.\footnote{More precisely, this field is an ``adjoint" of a $U(1)$ node, i.e. it is actually a singlet.} This corresponds to entering the Higgs branch by sliding out of the plane the segment between nodes -1 and 0 in \fref{Web_2_Hypers_Phase2_Mutated}. Plugging this VEV into the superpotential \eref{W_quiver_tail_1} makes the pair of bifundamental fields extended between nodes -1 and 0 massive, effectively decoupling node 0 from the rest of the quiver. This operation corresponds to sliding the 5-brane segment that stretches between the 7-branes -1 and the 0 along them to infinity. The superpotential becomes
\begin{equation}
\label{eq:Wtailstripped}
W=M_{1,3}Y_{3,-1}Y_{-1,1}+M_{2,3}Y_{3,-1}Y_{-1,2}+M_{1,4}Y_{4,-1}Y_{-1,1}+M_{2,4}Y_{4,-1}Y_{-1,2}\,.
\end{equation}

\begin{figure}[h!]
\centering
\includegraphics[height=6cm]{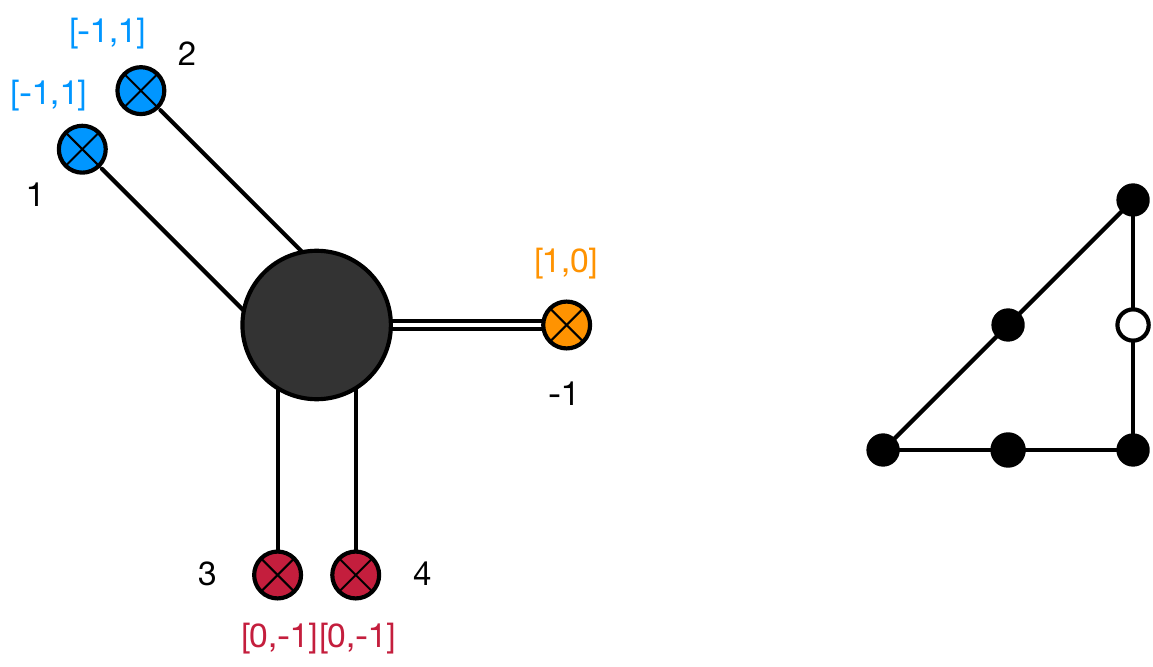}
\caption{Brane web obtained from \fref{Web_2_Hypers_Phase2_Mutated} by decoupling brane 0 and the corresponding GTP.}
\label{Web_2_Hypers_Phase2_GTP}
\end{figure}

\fref{Web_2_Hypers_Phase2_GTP} shows the web after decoupling the tail. It contains two 5-brane legs that terminate on a single 7-brane, i.e. it corresponds to a GTP with a white dot, as shown in the same figure. Then, it should be possible to alternatively construct the corresponding twin quiver using the algorithm discussed Section \sref{section_constructing_twin_quivers} including its third step, i.e. node merging. It is straightforward to verify that merging nodes -1 and 0 of the quiver in \fref{T2} into a single rank-2 node produces the desired quiver. Implementing the identification of nodes in the superpotential results in
\begin{eqnarray}
\label{Wmerged}
W&=&X_{1,4}(\tilde{X}_{4,-1}\tilde{X}_{-1,1}+X_{4,-1}X_{-1,1})+X_{2,3}\tilde{X}_{3,-1}(\tilde{X}_{-1,2}  +  X_{3,-1}X_{-1,2}) \\ \nonumber 
& + & X_{1,3}(\tilde{X}_{3,-1}\tilde{X}_{-1,1}+X_{3,-1}X_{-1,1})+X_{2,4}(\tilde{X}_{4,-1}\tilde{X}_{-1,2}+X_{4,-1}X_{-1,2})\,.
\end{eqnarray}
where we have assumed generic coefficients for each monomial and combined the bifundamentals connected to the merged nodes into doublets. Upon obvious relabeling, both the quiver and the superpotential \eqref{Wmerged} agree with the ones we previously derived by removing the tail.

\section{Additional examples}

\label{section_additional_examples}

In Section \sref{section_multiple_toric_phases}, we presented evidence that the different toric phases of the original theory allow to explore different chambers of the extended Coulomb branch of the $5d$ theory. This section contains additional examples that go beyond free theories, showing that this is a general phenomenon.

\subsection{Revisiting the free hypermultiplet: self-dual webs}

\begin{figure}[h!]
\centering
\includegraphics[height=6cm]{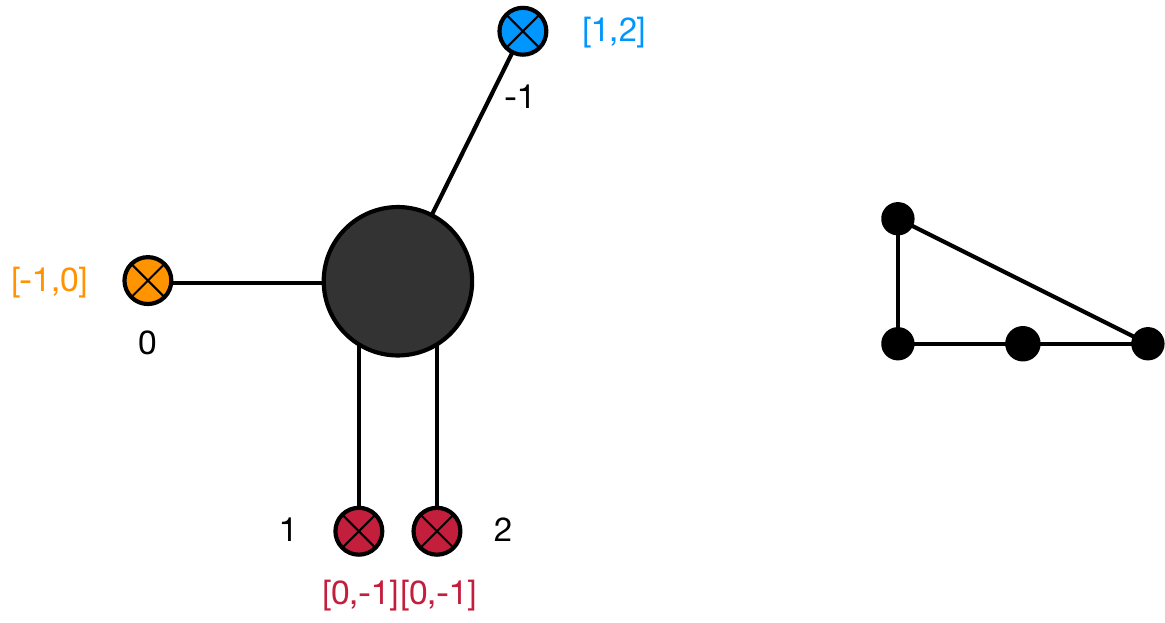}
\caption{Brane web for ``$SU(1)_1$", i.e. a free hypermultiplet. It corresponds to $\mathbb{C}^2/\mathbb{Z}_2 \times \mathbb{C}$.}
\label{web_SU1_1}
\end{figure}

When the class of webs in \fref{web_ZN_conifold} is specialized for $N=1$ as in \fref{1_hyper}, it engineers a single free hypermultiplet in $5d$. Formally, this theory can be regarded as ``$SU(1)_0$". The web for ``$SU(1)_1$", shown in \fref{web_SU1_1}, provides an alternative realization of a free hypermultiplet.

\begin{figure}[h!]
\centering
\includegraphics[height=6.5cm]{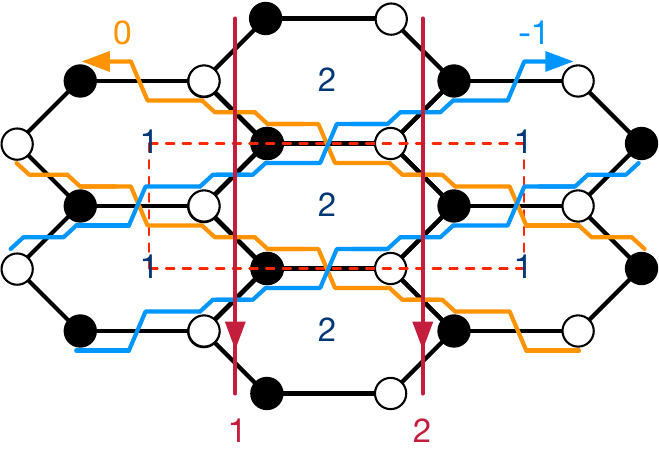}
\caption{Brane tiling for $\mathbb{C}^2/\mathbb{Z}_2 \times \mathbb{C}$. We show the zig-zag paths associated to the legs of the web in \fref{web_SU1_1}.}
\label{tiling_C2Z2xC}
\end{figure}

This web is dual to the $\mathbb{C}^2/\mathbb{Z}_2 \times \mathbb{C}$ singularity. There is a single toric phase for D3-branes proving this geometry, whose brane tiling is shown in \fref{tiling_C2Z2xC}.

\begin{figure}[h!]
\centering
\includegraphics[height=3.5cm]{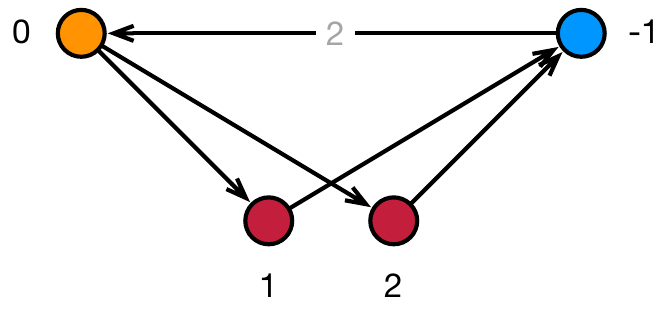}
\caption{Twin quiver for \fref{web_SU1_1}.}
\label{SU1_1_quiver}
\end{figure}

Untwisting the brane tiling, we obtain the twin quiver shown in \fref{SU1_1_quiver}, whose superpotential is
\begin{equation}
    W=X_{1,-1}Y_{-1,0}X_{0,1}+X_{-1,0}X_{0,2}X_{2,-1} - X_{0,1}X_{1,-1}X_{-1,0} - Y_{-1,0}X_{0,2}X_{2,-1}\,.
\end{equation}

\begin{figure}[h!]
\centering
\includegraphics[height=6cm]{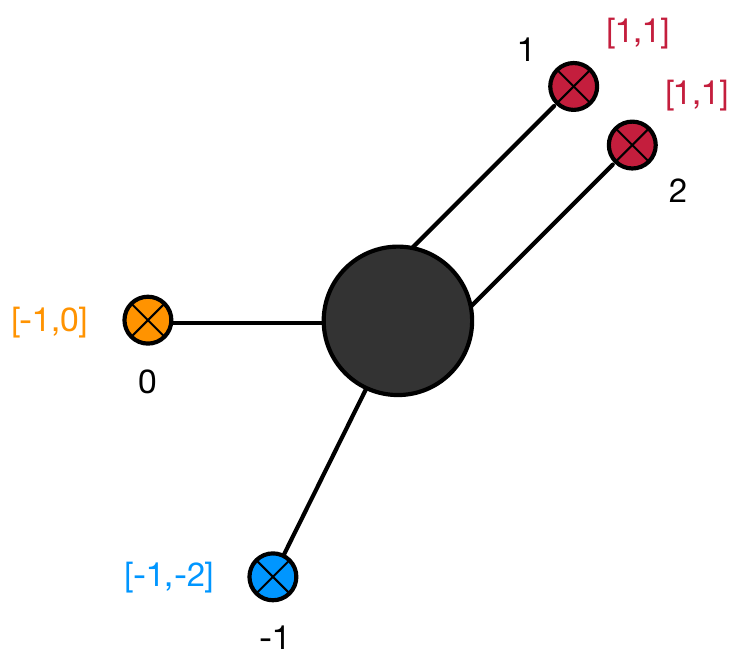}
\caption{Crossing the blue 7-brane in \ref{web_SU1_1}.}
\label{web_SU1_1_crossed}
\end{figure}

Crossing brane $-1$ in \fref{web_SU1_1} results in the web shown \fref{web_SU1_1_crossed}. Interestingly, the initial and mutated webs are $SL(2,\mathbb{Z})$ equivalent, therefore they describe the same $5d$ theory.

The observed self-duality is nicely captured by the twin quiver, which returns to itself when mutated on node -1 (up to relabeling nodes).

Finally, we note that upon crossing any of the red branes in \fref{web_SU1_1}, it decouples from the web. The same happens if one crosses any of the branes in \fref{1_hyper}. Both cases result in the same web, the one for $\mathbb{C}^3$. It is a straightforward to verify that the corresponding mutations of the twin quivers in figures \ref{SU1_1_quiver} and \ref{1_hyper} indeed result in the same quiver.

\subsection{The rank 1 $E_1$ theory}\label{E1}

\begin{figure}[h!]
\centering
\includegraphics[height=6cm]{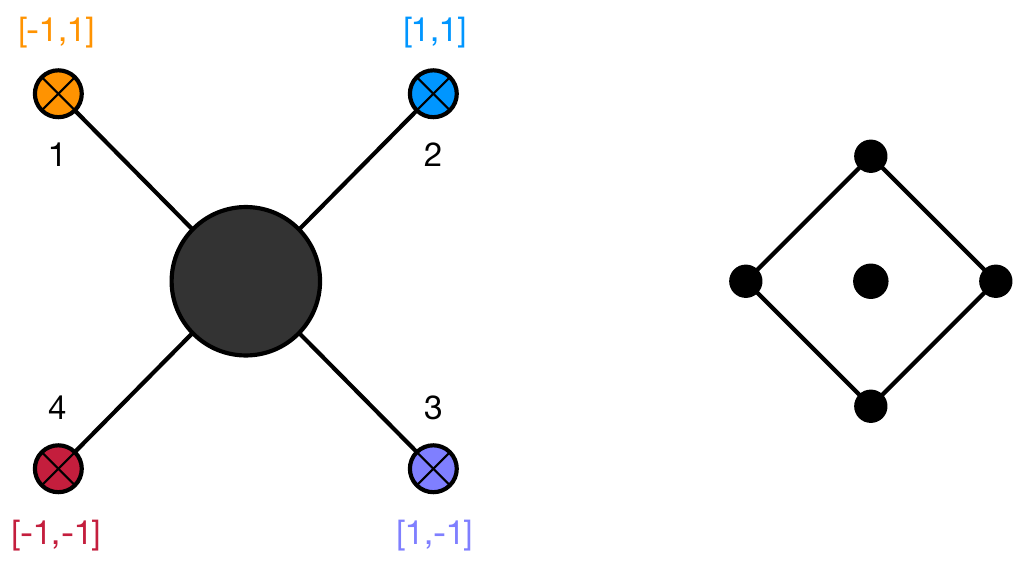}
\caption{Brane web for the $E_1$ theory. The dual toric diagram corresponds to $\mathbb{F}_0$.}
\label{web_E1}
\end{figure}

Let us explore a full-fledged interacting $5d$ theory, the $E_1$ theory, whose web is shown in \fref{web_E1}.

\begin{figure}[h!]
\centering
\includegraphics[height=6cm]{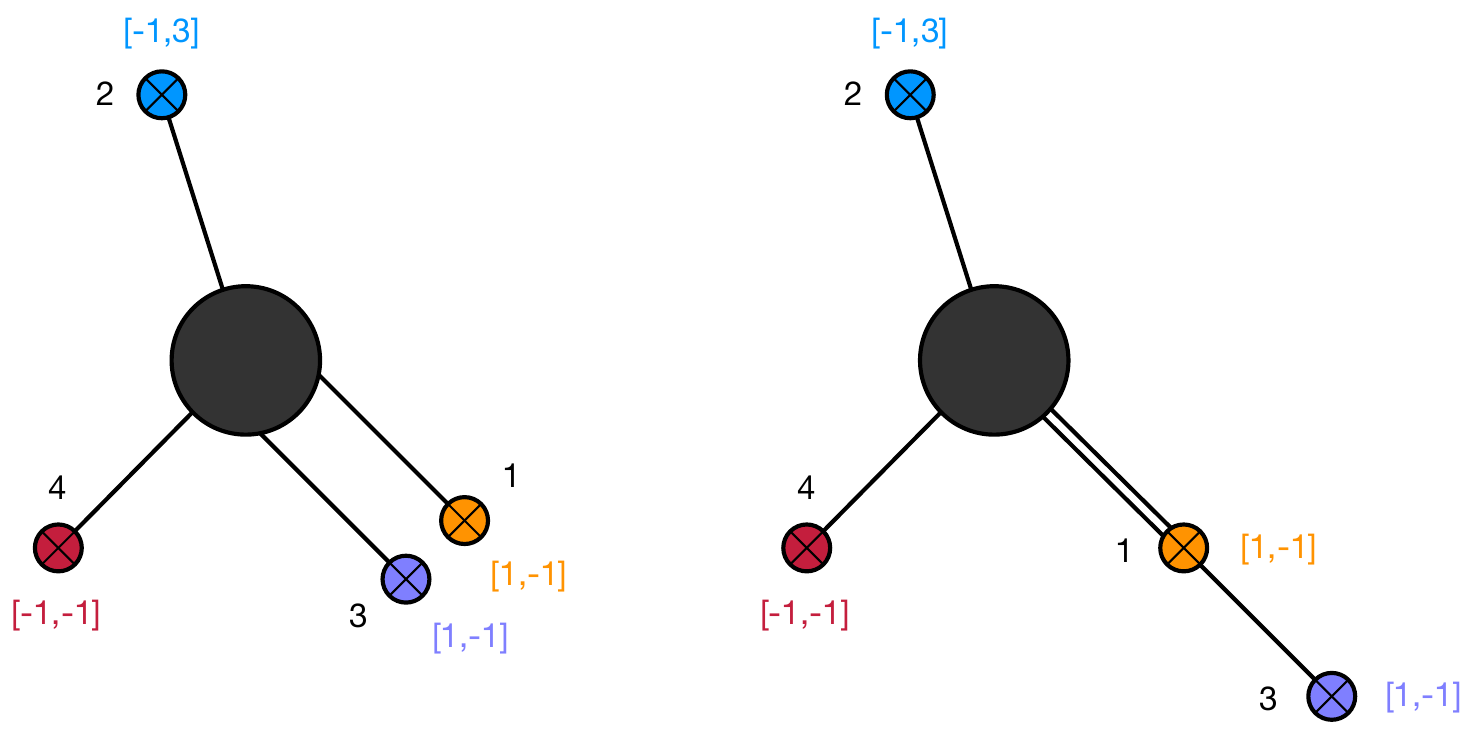}
\caption{Brane webs obtained by crossing 7-brane 1 in the $E_1$ theory.}
\label{web_E1_crossing}
\end{figure}

The four legs of the web are equivalent. Without loss of generality, let us consider crossing 7-brane 1 to the other side. Generically, we obtain the configuration on the left of \fref{web_E1_crossing}. However, in view of our previous experience, we expect the existence of another toric phase for the $E_1$ theory that results on a twin quiver associated to the root of the Higgs branch, shown on the right \fref{web_E1_crossing}.

\begin{figure}[h!]
\centering
\includegraphics[height=6cm]{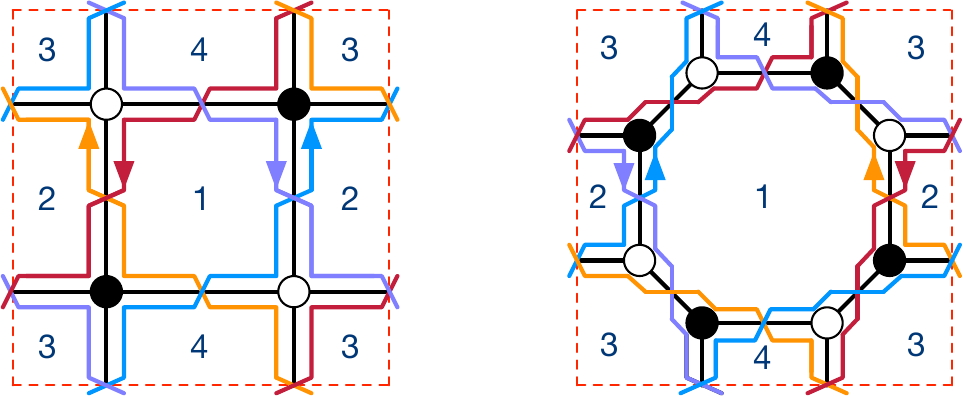}
\caption{Brane tilings for the two toric phases of $\mathbb{F}_0$. We distinguish the zig-zags with the colors of the corresponding 7-branes in \fref{web_E1}.} 
\label{tilings_F0}
\end{figure}

As shown in \fref{web_E1} the web for $E_1$ is dual to the toric diagram for $\mathbb{F}_0$. There are two toric phases for D3-branes probing this geometry \cite{Feng:2001xr,Feng:2001bn,Feng:2002zw,Franco:2005rj}. We present the corresponding brane tilings and quivers in Figure \ref{tilings_F0}.

\begin{figure}[h!]
\centering
\includegraphics[height=6cm]{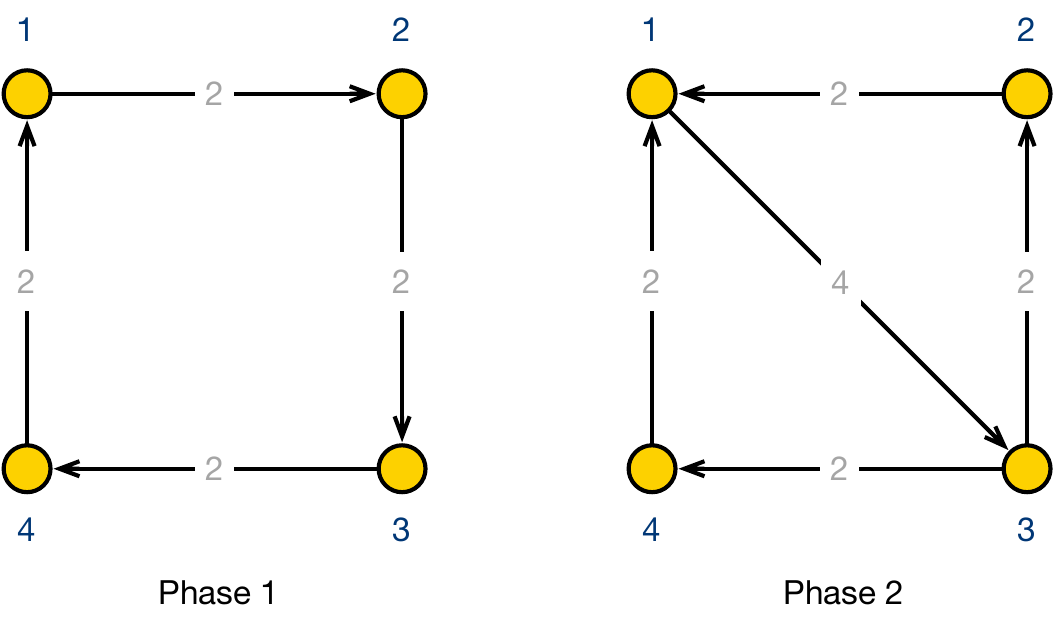}
\caption{Original quivers for the two toric phases of $\mathbb{F}_0$.}
\label{quivers_F0}
\end{figure}

\begin{figure}[h!]
\centering
\includegraphics[height=6cm]{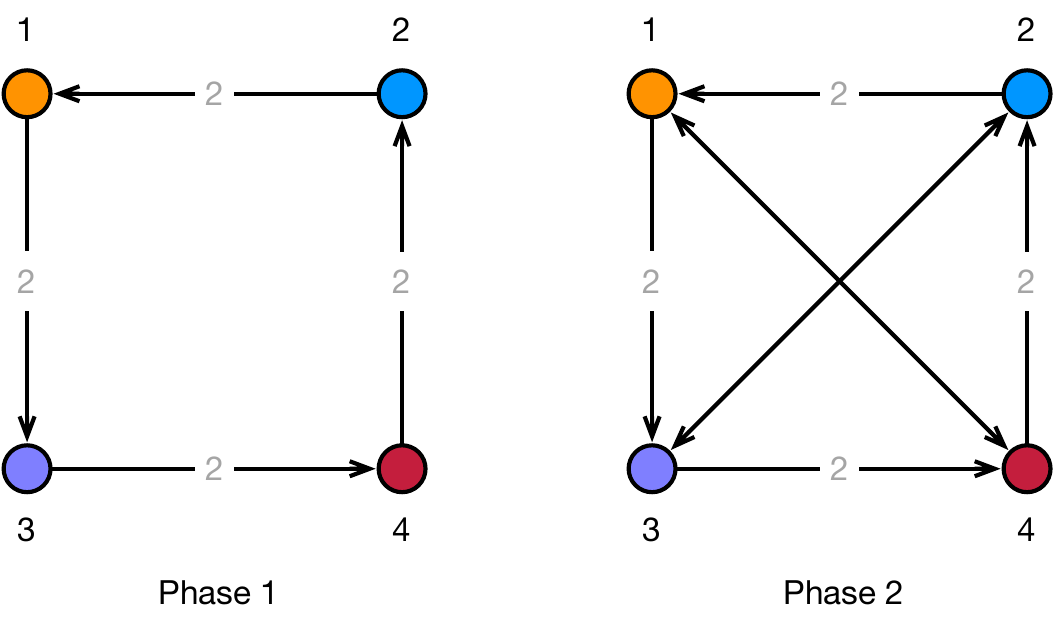}
\caption{Twin quivers obtained from the two toric phases of $\mathbb{F}_0$.} 
\label{twin_quivers_F0}
\end{figure}

Untwisting the brane tilings in \fref{tilings_F0}, we obtain the twin quivers in \fref{twin_quivers_F0} (see \cite{Feng:2005gw} for further details).

\begin{figure}[h!]
\centering
\includegraphics[height=5cm]{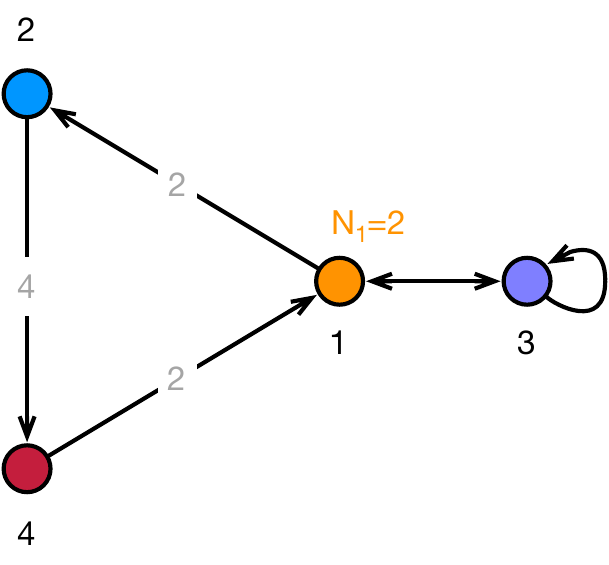}
\caption{Mutating the twin quiver for phase 2 in \fref{twin_quivers_F0} gives rise to a quiver tail. Here we show a mutation on node 1.}
\label{quiver_E1_tail}
\end{figure}

As expected, the two twin quivers differ by the presence of bidirectional arrows. Their superpotentials can be easily determined from the corresponding graphs. The extra arrows turn otherwise toric nodes, i.e. nodes with $N_f=2N_c$, into non-toric ones. This agrees with the expectation from \fref{web_E1_crossing}. Indeed, while it is straightforward to check that mutating any node -- they are all equivalent -- of phase 1 in \fref{twin_quivers_F0} gives a quiver naturally associated to the web on the left of \fref{web_E1_crossing}, mutating any node of phase 2 in \fref{twin_quivers_F0} gives \fref{quiver_E1_tail}. The triangular sub-quiver consisting of the yellow, red and light blue nodes agrees with the corresponding sub-web on the right of  \fref{web_E1_crossing}. In addition, the tail of the twin quiver agrees with the web.

Similarly to the example in Section \sref{TailDecouplingAndNodeMerging}, one can check that node merging in the mutation of phase 1 produces the same result as tail removal in the mutation of phase 2.

\subsection{The $E_2$ theory}\label{dP2}

\begin{figure}[h!]
\centering
\includegraphics[height=6cm]{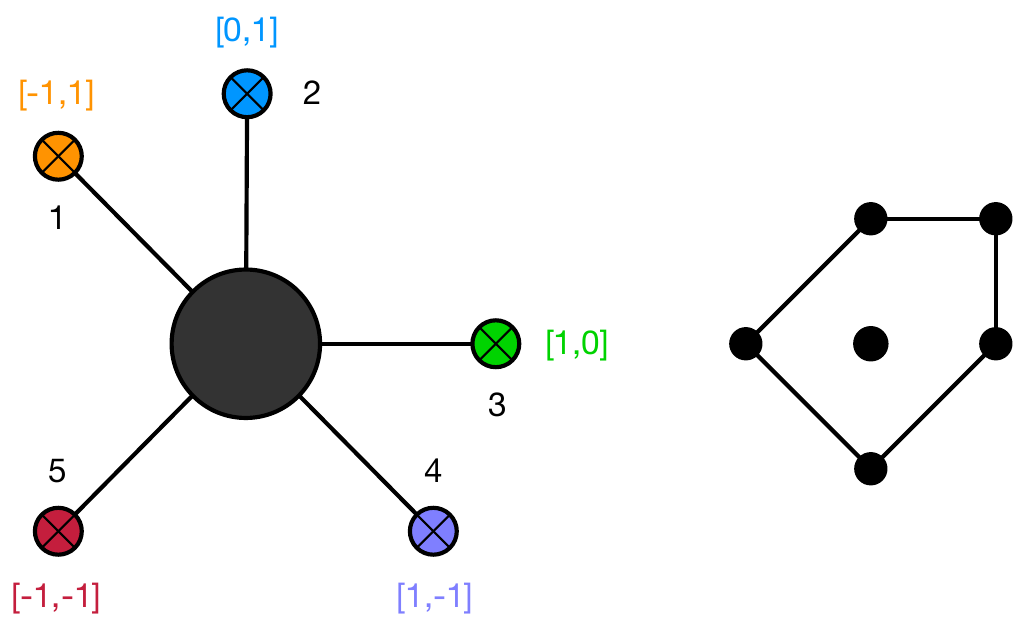}
\caption{Brane web for the $E_2$ theory. The dual toric diagram corresponds to $dP_2$.}
\label{web_E2}
\end{figure}

Let us now turn our attention to the $E_2$ theory, whose brane web is shown in \fref{web_E2}.

\begin{figure}[h!]
\centering
\includegraphics[height=6cm]{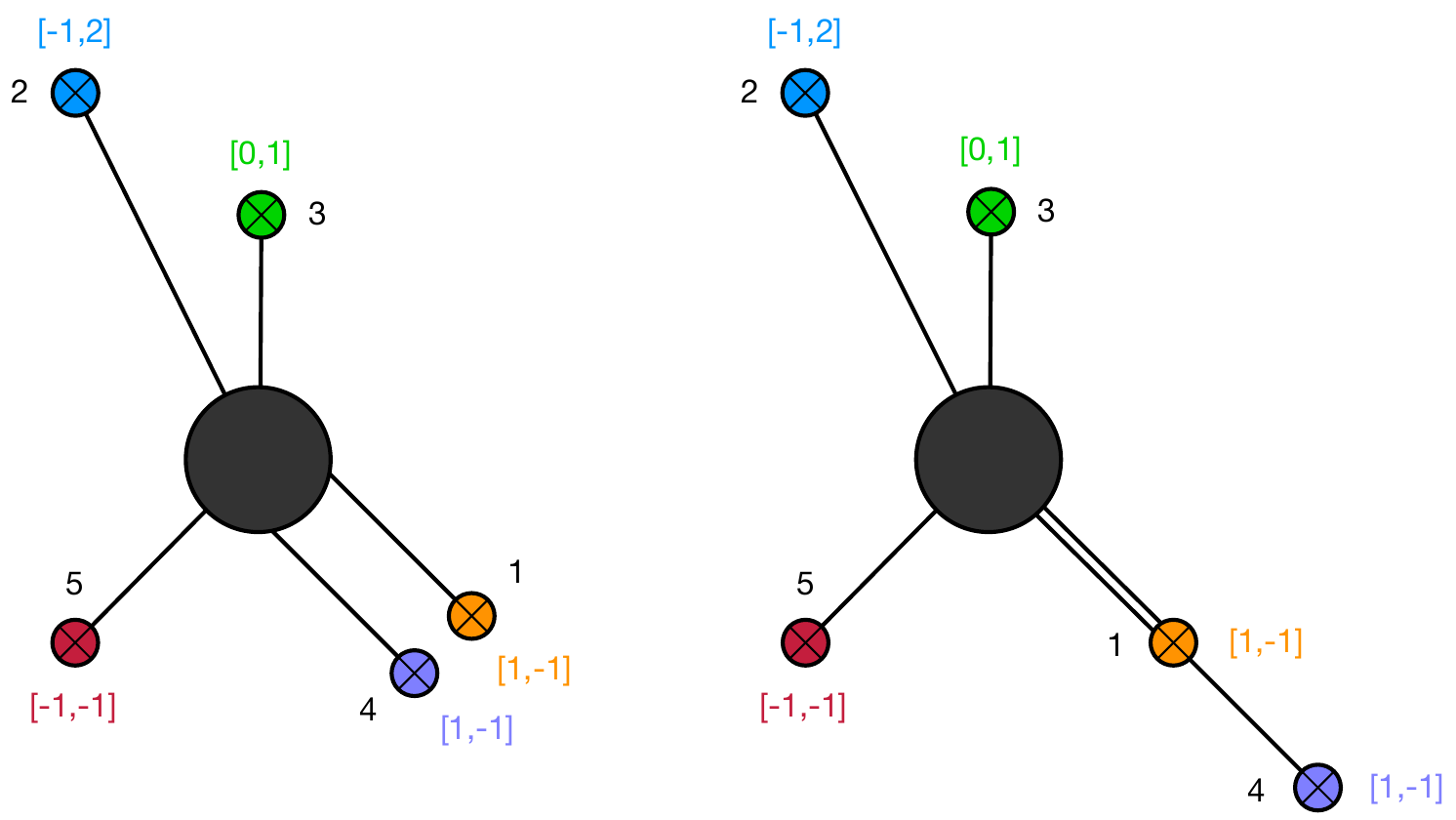}
\caption{Brane webs obtained by crossing 7-brane 1 in the $E_2$ theory.}
\label{web_E2_crossing}
\end{figure}

Crossing the 7-brane 1 we obtain, generically, the web shown on the left of \fref{web_E2_crossing}. Once again, there is a root of the Higgs branch in the extended Coulomb branch, whose web is shown on the right of \fref{web_E2_crossing}.

\begin{figure}[h!]
\centering
\includegraphics[height=6cm]{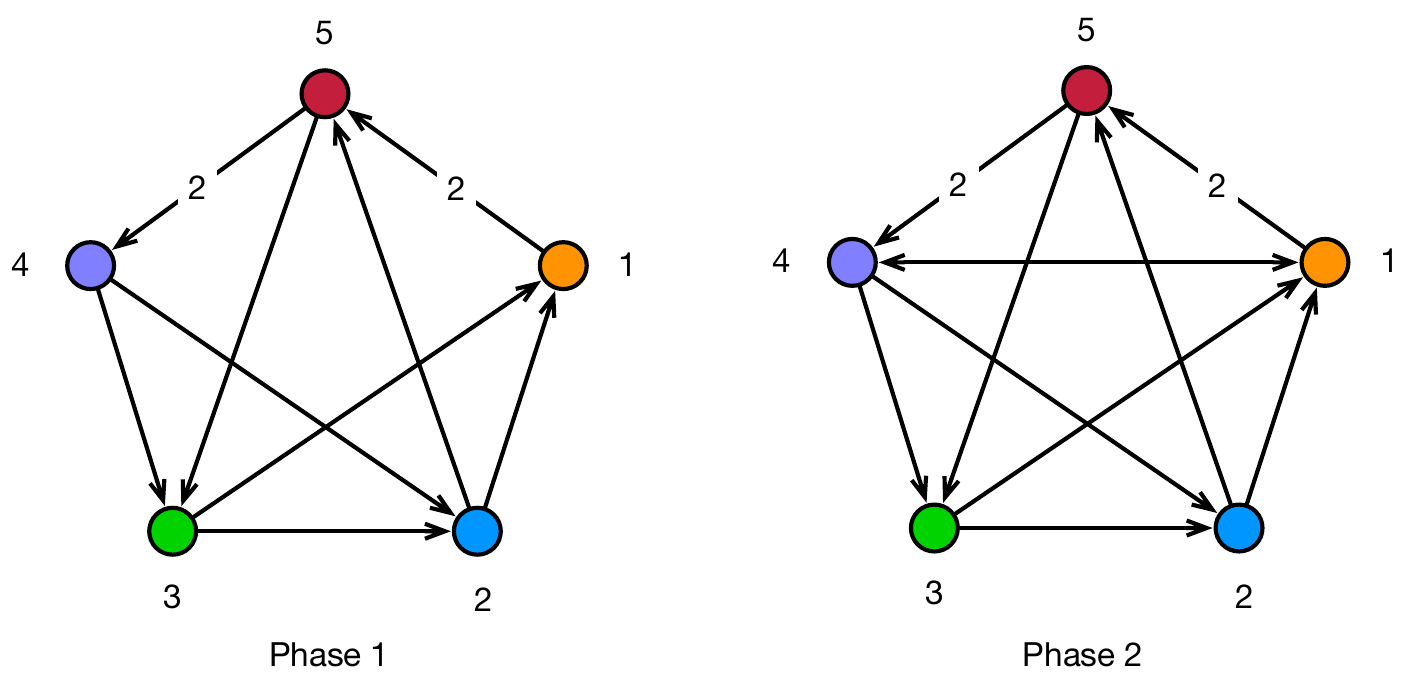}
\caption{The two twin quivers for the $E_2$ theory, which come from the two toric phases of $dP_2$.}
\label{twin_quivers_dP2}
\end{figure}

As shown in \fref{web_E2}, the $E_2$ web corresponds to the toric diagram for $dP_2$. There are two toric phases for D3-branes probing $dP_2$ \cite{Feng:2002zw}. As in previous examples, the webs in \fref{web_E2_crossing} can be related to these different phases. The brane tilings for the two toric phases of $dP_2$ and the BFTs obtained from them by untwisting were thoroughly discussed in \cite{Franco:2023flw}, to where we refer the interested reader for details. Here we just present the two associated twin quivers, which are shown in \fref{twin_quivers_dP2}. Once again, as anticipated, they only differ by bidirectional arrows, in this case stretched between nodes 1 and 4. As usual, it is straightforward to determine their superpotentials from the corresponding bipartite graphs.

\begin{figure}[h!]
\centering
\includegraphics[height=6cm]{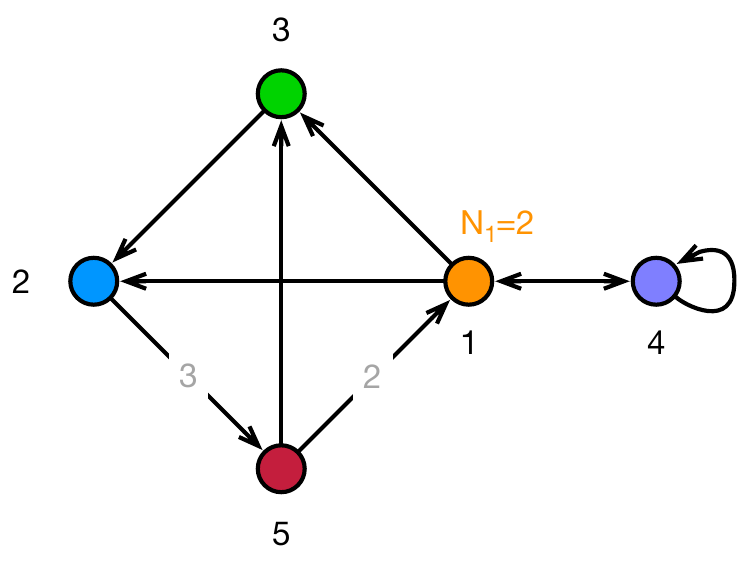}
\caption{Mutation of the twin quiver for phase 2 in \fref{twin_quivers_dP2} on node 1.}
\label{twin_quiver_dP2_tail}
\end{figure}

We are interested in studying how the crossing of 7-brane 1 in \fref{web_E2} translates in terms of twin quivers. Based on our previous experience, we expect that the web containing a tail in \fref{web_E2_crossing} correspond to the dualization of the twin quiver for phase 2 in \fref{twin_quivers_dP2} on node 1. Performing the dualization, we obtain the quiver in \fref{twin_quiver_dP2_tail}, which nicely shows the tail. Moreover, it is easy to verify that the rest of the quiver is consistent with the right web in \fref{web_E2_crossing}.

\subsection{A larger example}

\begin{figure}[h!]
\centering
\includegraphics[height=5cm]{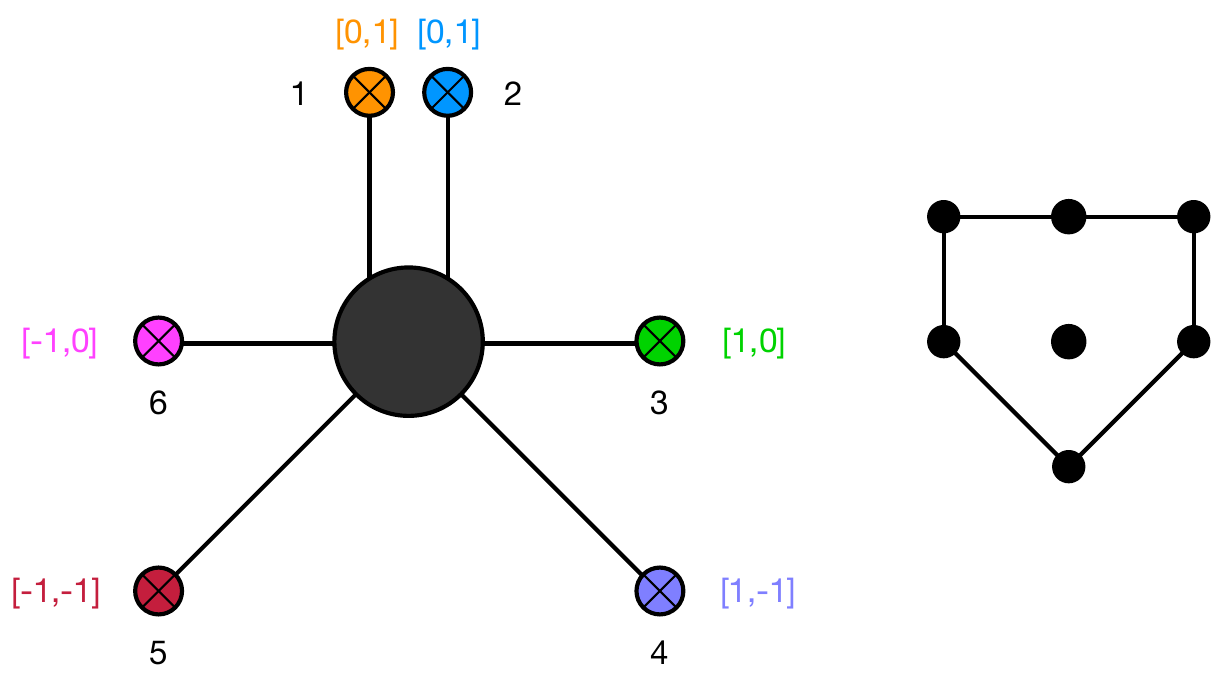}
\caption{A brane web and its dual toric diagram.}
\label{web_3}
\end{figure}

In this section we present an additional example illustrating our ideas. Consider the web and associated toric geometry shown in \fref{web_3}. This theory was thoroughly investigated in \cite{Franco:2023flw}, to where we refer the reader for further details.

\begin{figure}[h!]
\centering
\includegraphics[height=4.5cm]{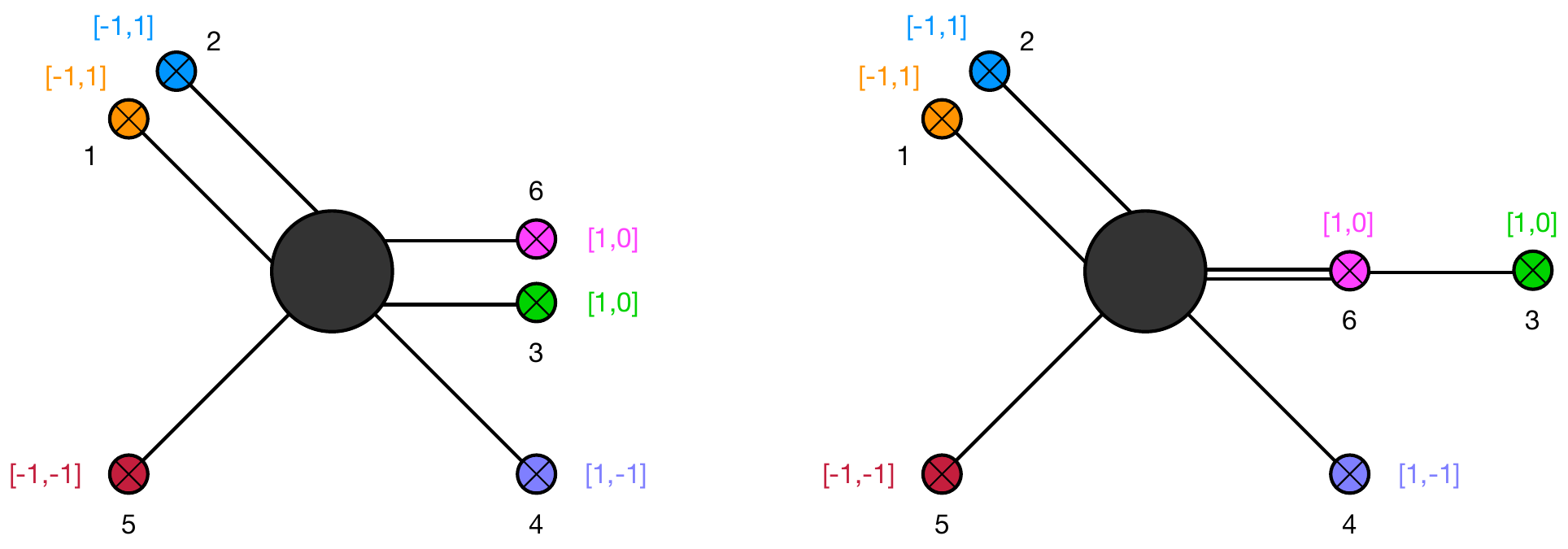}
\caption{Brane webs obtained by crossing 7-brane 6 in \fref{web_3}.}
\label{web_3_mutated}
\end{figure}

\fref{web_3_mutated} shows two webs obtained by crossing 7-brane 6.

\begin{figure}[h!]
\centering
\includegraphics[height=6cm]{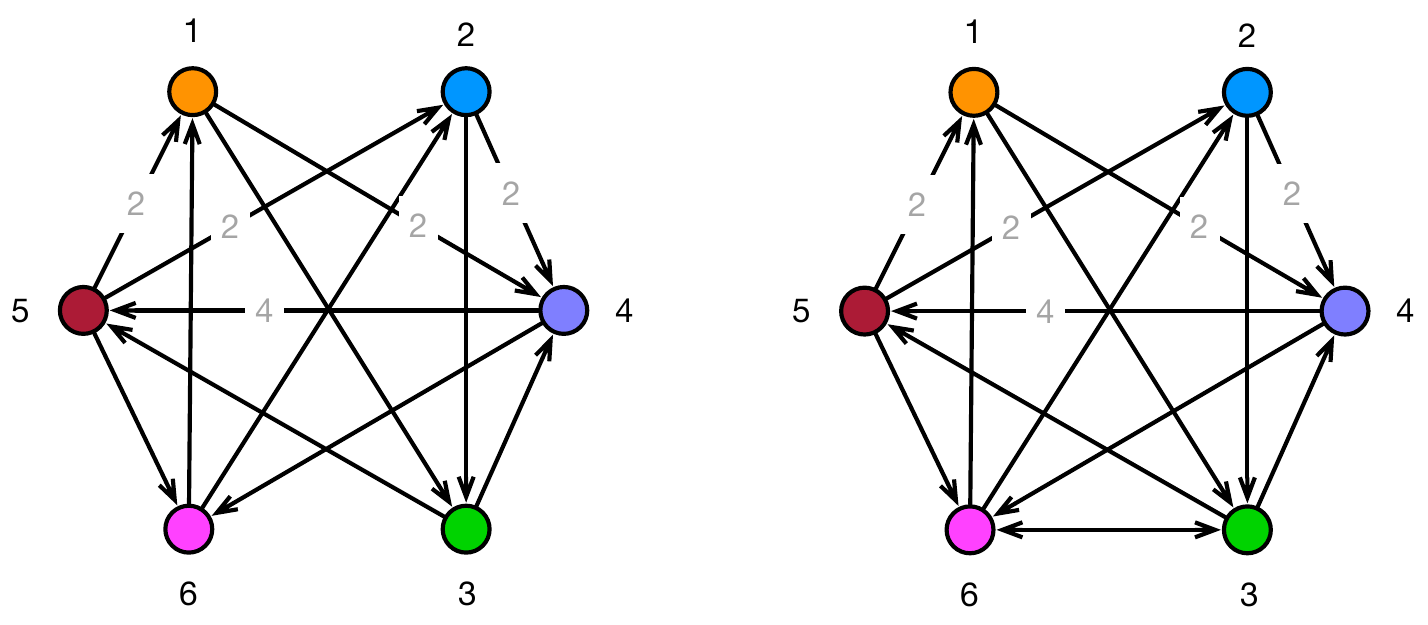}
\caption{Two twin quivers for the geometry in \fref{web_3}.}
\label{twin_quivers_web_3}
\end{figure}

Two twin quivers for the geometry in \fref{web_3} were presented in \cite{Franco:2023flw}.\footnote{While we have not carried out an exhaustive classification of toric phases for this geometry, it is reasonable to expect additional ones exist. We do not consider this question any further, since twin quivers associated to other phases are not relevant for our discussion.} We show them in \fref{twin_quivers_web_3}.\footnote{The superpotential for both theories can be determined from the bipartite graphs presented in \cite{Franco:2023flw}.} They differ by the presence of bidirectional arrows between nodes 3 and 6, which correspond to anti-parallel legs in \fref{web_3}.

\begin{figure}[h!]
\centering
\includegraphics[height=6cm]{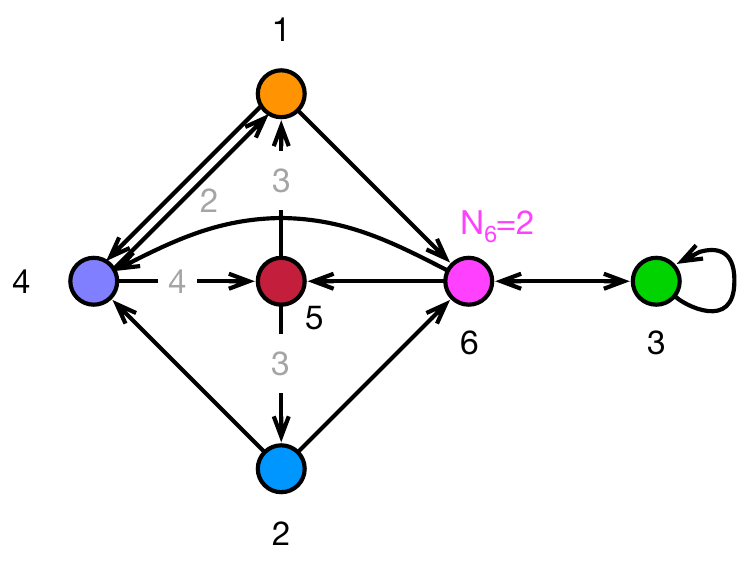}
\caption{Mutation of the twin quiver on the right of \fref{twin_quivers_web_3} on node 6.}
\label{twin_quiver_web_3_tail}
\end{figure}

To obtain the quiver describing the web with a tail on the right of \fref{web_3}, we consider the mutation of the quiver on the right of \fref{twin_quivers_web_3} on node 6. The resulting quiver, shown in \fref{twin_quiver_web_3_tail} exhibits the expected tail.

\section{Tails as building blocks}

\label{section_tails_as_building_blocks}

\begin{figure}[h!]
\centering
\includegraphics[height=4cm]{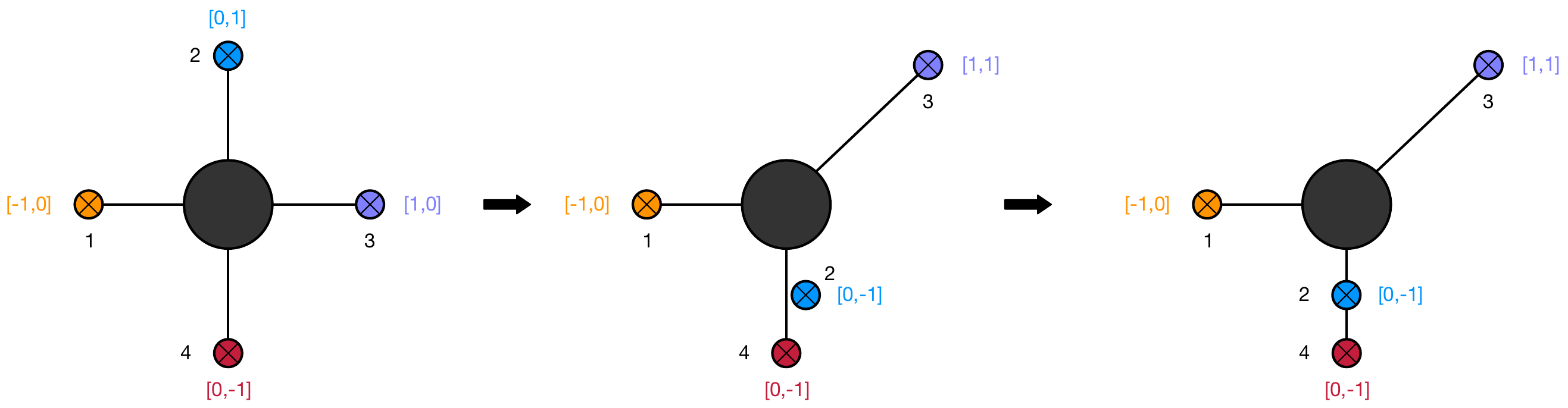}
\caption{Web for 1 free hyper and its Higgs branch.}
\label{web_tail_gluing}
\end{figure}

We now discuss a puzzle that arises in the quiver description of certain brane configurations and propose a solution. The problem is rather generic but, for concreteness, we illustrate it with an example. Consider the web for a free hypermultiplet shown in \fref{web_tail_gluing}.\footnote{This is the same web in \fref{1_hyper}. For simplicity, we have labeled the 7-branes differently.} Let us now cross one of the 7-branes. Without loss of generality, we can assume it is brane 2. In this process, brane 2 decouples from the rest of the web and can be moved away to infinity. This process is nicely captured by the twin quiver in \fref{T2}, from which the corresponding node disappears after mutating it. However, as shown in \fref{web_tail_gluing}, we can deal with brane 2 in a different way. Instead of sending it to infinity, we move it on top of the external leg that terminates on brane 4, splitting it. We recognize the resulting configuration as a tail. 

The previous discussion raises a puzzle. The starting web in \fref{web_tail_gluing} corresponds to the conifold, which has a single toric phase. This results in a unique twin quiver, the one in \fref{1_hyper}. Contrary to previous examples, there is no extra toric phase from which we could obtain an alternative twin quiver by untwisting that, in turn, would give rise to the tail when the node associated to brane 2 is mutated. In short, there does not seem to be a twin quiver capable of reproducing the last step in \fref{web_tail_gluing}.

\begin{figure}[h!]
\centering
\includegraphics[height=5cm]{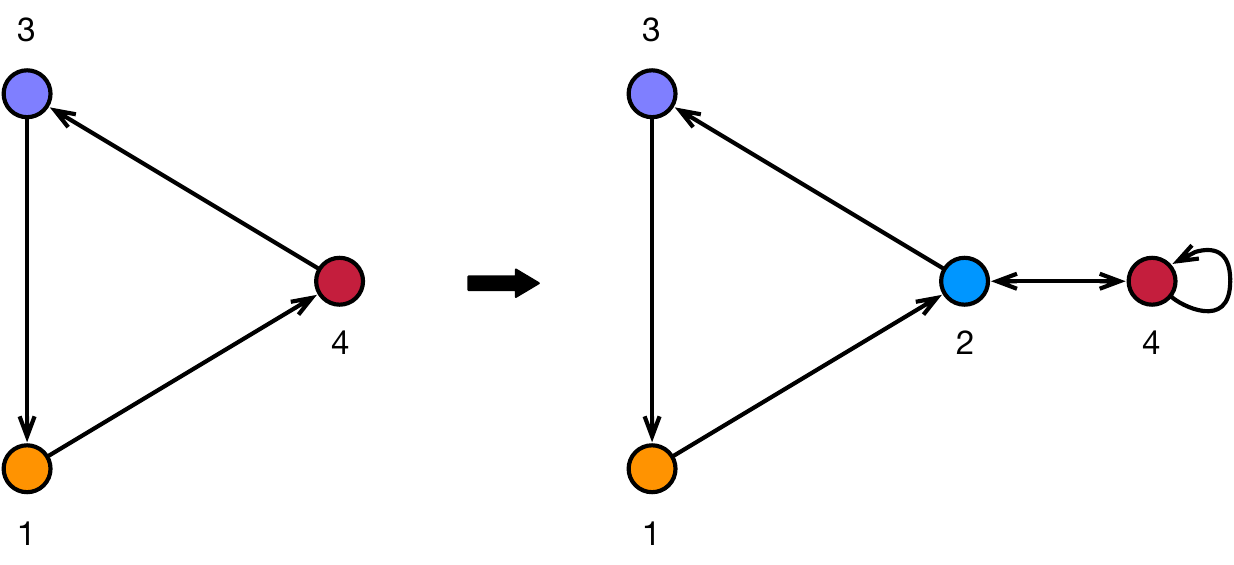}
\caption{Left: the twin quiver associated to the connected component of the web at the center of \fref{web_tail_gluing}. Right: we propose the twin quiver for the web on the right pf \fref{web_tail_gluing} is obtained by appending a tail.}
\label{quiver_tail_gluing}
\end{figure}

We propose to construct the corresponding twin quiver by appropriately appending a tail, as shown in \fref{quiver_tail_gluing}. In this case, in accordance with the web in \fref{web_tail_gluing}, the rank of node 2 at the ``root" of the quiver tail is 1. This quiver is certainly consistent with the basic properties of the web in \fref{web_tail_gluing}, but what about the superpotential? Based on the symmetries of the configuration, it is natural to conjecture that
\beq
    W=X_{4,4}X_{4,2}X_{2,4}+X_{2,3}X_{3,1}X_{1,2}X_{2,4}X_{4,2}\,.
\eeq

We can test this proposal by mutating the quiver on node 2, which corresponds to taking brane 2 back to its original position. The resulting quiver is given in \fref{quiver_tail_gluing_mutated} below, and its superpotential is
\beq
W=X_{1,3}X_{3,2}X_{2,1}+X_{3,1}X_{1,4}X_{4,3}+X_{2,4}X_{4,3}X_{3,2}+X_{4,2}X_{2,1}X_{1,4}\,.
\eeq
This result supports our proposal, since we obtained a natural extension of \fref{1_hyper} by the addition of bidirectional arrows, with a superpotential that respects the symmetry of the web on the left of \fref{web_tail_gluing}.

\begin{figure}[h!]
\centering
\includegraphics[height=6cm]{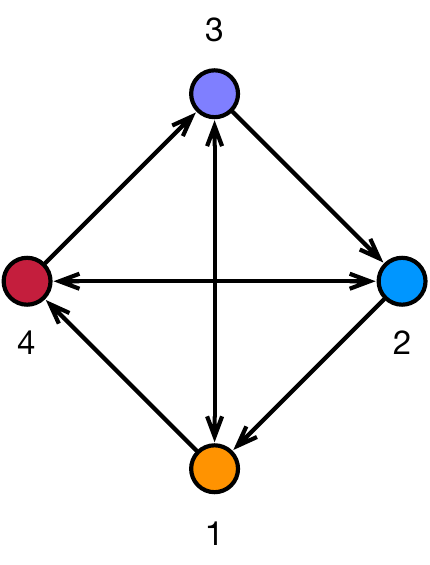}
\caption{Mutation of the quiver on the right of \fref{quiver_tail_gluing} on node 2.}
\label{quiver_tail_gluing_mutated}
\end{figure}

\section{Conclusions} 

\label{section_conclusions}

Twin quivers were introduced in \cite{Franco:2023flw} as new tools for studying the general class of $5d$ SCFTs encoded in webs of 5- and 7-branes or, equivalently, GTPs. In this paper we investigated the non-uniqueness of twin quivers that follows from the multiplicity of toric phases for a given toric Calabi-Yau 3-fold used in their construction. This phenomenon exists even for brane webs associated to ordinary toric diagrams, i.e. it is independent from the presence of white dots in GTPs. As noted in \cite{Franco:2023flw}, twin quivers originating from different toric phases differ by bidirectional arrows. We found that such bidirectional arrows modify the mutation of twin quivers, in particular giving rise to quiver tails. We observed that quiver tails can describe certain sub-configurations of the brane webs obtained via Hanany-Witten transitions, which arise at roots of the Higgs branch in the extended Coulomb branch of the $5d$ theories. We offered evidence of the appearance of tails in various examples and explored the internal consistency of the emerging picture. Below we discuss some directions for future research.

We have seen that, upon mutation, twin quivers with single bidirectional arrows between pairs of nodes naturally give rise to the tail structure associated to the corresponding webs. However, generically, there exist toric phases that result in twin quivers with more than one bidirectional arrow connecting a given pair of nodes. It would be interesting to clarify the web interpretation of the mutations of such quivers and whether they are in correspondence with all the roots of the Higgs branch on the extended Coulomb branch. From this perspective, the tail decoupling of Section \sref{TailDecouplingAndNodeMerging} would naturally correspond to entering a Higgs branch of the theory. Note that predicting the number of toric phases for a given toric CY$_3$ is an open question. Our work suggests an interesting connection between this problem and the classification of roots of the Higgs branch in the extended Coulomb branch of $5d$ theories or, equivalently, of the corresponding brane webs. It would be interesting to explore this correspondence in further detail and determine whether it sheds light on the enumeration of toric phases. 

Another question worth investigating is the ``surgery" of twin quivers, refining or extending our gluing/removing procedure introduced in Sections \sref{section_tails_as_building_blocks} and \sref{TailDecouplingAndNodeMerging} as a method for generating twin quivers for GTPs. Related to this, we have seen that tail decoupling can be used to derive the twin quiver associated to a GTP with white dots. It is thus very natural to ask whether there is an interpretation of this procedure at the level of the original brane tiling. Likewise, one could start with a web with no white dot in the GTP and, upon mutation, find a web described by a GTP with white dots\footnote{Take for instance the $+_{N,1}$ theory of Section \sref{sect:conifold/Zn}: for $N>2$, the analog of our $-1$ mutation would produce a web encoded in a GTP with white dots.} which would naturally lead to the same question. This could provide a hint towards the more ambitious goal of finding a generalization of brane tilings applicable to generic GTPs. It is expected that such a generalization would encode the BPS quivers of the corresponding $5d$ theories.

Finally, it would be interesting to carry out a detailed study of the moduli spaces of the twin theories and how they depend on the presence of bidirectional arrows, even in the simple case of GTPs without white dots. It is tempting to speculate that bidirectional arrows introduce extra branches in the moduli space of the twin theory related to the tails that appears upon mutation. Moreover, their excision as in \sref{TailDecouplingAndNodeMerging}, could be mapped to entering a Higgs branch and give rise to the the twin moduli space for GTPs with white dots. In this respect, it would be very interesting to make contact with the description of the geometries associated to GTPs with white dots presented in \cite{Bourget:2023wlb}. We plan to come back to these issues in the future.

\acknowledgments

We would like to thank Guillermo Arias-Tamargo, Francesco Benini, Sergio Benvenuti, Antoine Bourget, Julius Grimminger and Amihay Hanany for discussions. S.F. would like to thank Rak-Kyeong Seong for previous collaboration on related topics. We would like to acknowledge the Simons Physics Summer Workshop and the Simons Center for Geometry and Physics for hospitality during part of this work. S.F. is supported by the U.S. National Science Foundation grants PHY-2112729 and DMS-1854179.  D.R.G is supported by Spanish national grant MCIU-22-PID2021-123021NB-I00 as well as the Principado de Asturias grant SV-PA-21-AYUD/2021/52177.


\bibliographystyle{JHEP}
\bibliography{mybib}

\end{document}